\documentclass[preprint,preprintnumbers,amsmath,amssymb,nofootinbib,superscriptaddress]{revtex4}

\usepackage{epsfig,latexsym,cancel,amssymb,amsmath}
\usepackage{color}
\usepackage{graphicx}

\newcommand{\beq}{\begin{eqnarray}}
\newcommand{\eeq}{\end{eqnarray}}

\newcommand{\ba}{\begin{array}}
\newcommand{\ea}{\end{array}}

\newcommand{\CPv}{{\textrm{\fontsize{6}{11}\selectfont CP}\!\!\!\!\!\!\!\diagup}}

\long\def\symbolfootnote[#1]#2{\begingroup%
\def\thefootnote{\fnsymbol{footnote}}\footnote[#1]{#2}\endgroup}

\newcommand{\eq}[1]{Eq.~\eqref{#1}}
\newcommand{\eqs}[2]{Eqs.~\eqref{#1} and \eqref{#2}}

\newcommand{\fig}[1]{Fig.~\ref{#1}}

\newcommand{\no}{\nonumber}
\def\lsim{\mathrel{\rlap{\lower4pt\hbox{\hskip1pt$\sim$}}
    \raise1pt\hbox{$<$}}}         
\def\gsim{\mathrel{\rlap{\lower4pt\hbox{\hskip1pt$\sim$}}
    \raise1pt\hbox{$>$}}}         

\begin{document}

\preprint{PI/UAN-2009-407FT}

\title{CP violation Beyond the MSSM:\\ Baryogenesis and Electric Dipole Moments}

\author{Kfir Blum}
\affiliation{Department of Particle Physics and Astrophysics, Weizmann
  Institute of Science, Rehovot 76100, Israel}
\author{Cedric Delaunay}
\affiliation{Department of Particle Physics and Astrophysics, Weizmann
  Institute of Science, Rehovot 76100, Israel}
\author{Marta Losada}
\affiliation{Centro de Investigaciones, Cra 3 Este No 47A-15,
  Universidad Antonio Nari\~{n}o, Bogot\'a, Colombia}
\author{Yosef Nir}
\affiliation{Department of Particle Physics and Astrophysics, Weizmann
  Institute of Science, Rehovot 76100, Israel}
\author{Sean Tulin}
\affiliation{Theory Group, TRIUMF, 4004 Wesbrook Mall, Vancouver, BC,
  V6T 2A3, Canada}

\date{\today}

\begin{abstract}
  We study electroweak baryogenesis and electric dipole moments in the
  presence of the two leading-order, non-renormalizable operators in
  the Higgs sector of the MSSM.  Significant qualitative and
  quantitative differences from MSSM baryogenesis arise due to the
  presence of new CP-violating phases and to the relaxation of
  constraints on the supersymmetric spectrum (in particular, both
  stops can be light).  We find: (1) spontaneous baryogenesis, driven
  by a change in the phase of the Higgs vevs across the bubble wall,
  becomes possible; (2) the top and stop CP-violating sources can
  become effective; (3) baryogenesis is viable in larger parts of
  parameter space, alleviating the well-known fine-tuning associated
  with MSSM baryogenesis.  Nevertheless, electric dipole moments
  should be measured if experimental sensitivities are improved by about one
  order of magnitude.
\end{abstract}

\maketitle


\section{Introduction}
\label{sec:intro}
Electroweak baryogenesis (EWBG) is an attractive mechanism for
generating the baryon asymmetry of the Universe (BAU).  Its primary
attraction is the possibility to experimentally test two of the three
Sakharov conditions.  Aspects of the departure from thermal
equilibrium via the electroweak phase transition (EWPT) can be
explored in collider experiments, while CP violation can be tested in
electric dipole moment (EDM) searches.

In the EWBG picture, electroweak symmetry breaking proceeds via a
first-order phase transition, where bubbles of broken SU(2)$_L$
symmetry nucleate and expand in a background of unbroken symmetry.
CP-violating interactions within the bubble walls lead to the
production of CP-asymmetric charge density of left-handed fermions.
This charge, diffusing ahead of the wall into the unbroken phase, is
converted into the BAU by non-perturbative, electroweak sphaleron
processes.  To the extent that electroweak sphalerons are inactive
after electroweak symmetry breaking, the baryon density ``freezes
out'' once it is captured by the advancing bubble wall.  This
mechanism satisfies the Sakharov criteria \cite{Sakharov:1967dj} and
generates the BAU provided two conditions are met: (1) the phase
transition is ``strongly'' first-order (otherwise electroweak
sphalerons are active within the broken phase and washout the BAU),
and (2) the CP violation is sufficient to generate the observed BAU.
Neither of these conditions are met in the Standard Model
(SM)~\cite{sm}.

Beyond the SM, the most widely studied EWBG model has been the Minimal
Supersymmetric Standard Model (MSSM).  However, there are a series of
tensions that make this scenario severely constrained by experiment.
First, there is tension in the top squark (``stop'') sector.  A strong
first-order phase transition requires at least one light stop (which
must be mostly $\widetilde t_R$, to avoid large contributions to the
$\rho$ parameter and due to null searches for a light
sbottom~\cite{Drees:1990dx,Amsler:2008zzb}).  At the same time, the
large radiative corrections needed to push the Higgs boson mass above
the LEP bound $m_h>114$ GeV~\cite{Amsler:2008zzb} require that at
least one stop ($\widetilde t_L$) is very heavy \cite{Quiros:2000wk}.
Recently, the phase transition was studied in an effective theory with
a large stop hierarchy, concluding that successful EWBG is possible
only for $m_{\widetilde t_R} < 125$ GeV and $m_{\widetilde t_L} > 6.5$
TeV~\cite{Carena:2008vj}, rendering the scenario finely tuned.

Second, there is tension between having enough CP violation to produce
the BAU and evading stringent constraints from EDM searches. In the
MSSM, the CP-violating phases that drive EWBG arise in the
gaugino/higgsino sector. The same phases contribute to EDMs. While
one-loop contributions can be sufficiently suppressed by making the
first two squark and slepton generations heavy, there exist two-loop
contributions that cannot be suppressed without spoiling EWBG
(assuming no fine-tuned cancellation between different EDM
contributions) and which predict a minimum value of the EDM. These
``irreducible'' EDMs strongly constrain the viable MSSM parameter
space: EWBG with universal gaugino phases is nearly ruled out. With
improvements by a factor $3\!-\!4$ in the upper bounds on the EDMs of
the electron or the neutron, MSSM baryogenesis will be possible only
in the so-called ``bino-driven'' scenario, where the CP-violating
phase associated with the U(1)$_Y$ gaugino is tuned to be much larger
than that of the SU(2)$_L$ gaugino~\cite{Li:2008ez,Cirigliano:2009yd}.

Third, there is tension in the mass of the pseudoscalar Higgs boson
$A_0$.  Large values of $m_A$ are preferred (i) to make the EWPT more
strongly first-order, and (ii) to evade constraints from $b \to s
\gamma$~\cite{Cirigliano:2009yd}.  However, the production of
left-handed charge during EWBG is enhanced when $m_A$ is light.  There
is also a tension in the value of $\tan\beta$ (i) from a compromise in
giving a large enough value of the Higgs mass versus a strong enough
phase transition, and (ii) from the constraints from $b \rightarrow s
\gamma$ for small values of $m_A$. All in all, from the theoretical
point of view, these tensions force the MSSM (if it is to account for
EWBG) into a narrow, finely tuned region of parameter space.

An attractive extension of the MSSM is the ``Beyond the MSSM'' (BMSSM)
scenario~\cite{Dine:2007xi}.  Here, a non-renormalizable contribution
to the MSSM superpotential is included
\beq\label{eq:BMSSM1}
W_{\textrm{BMSSM}} = W_{\textrm{MSSM}} +
\frac{\lambda}{M} \, (H_u \, H_d)^2 \; ,
\eeq
as well as a contribution to the soft SUSY-breaking Lagrangian
\beq\label{eq:BMSSM2}
\mathcal{L}_{\textrm{soft}}^{\textrm{BMSSM}} =
\mathcal{L}_{\textrm{soft}}^{\textrm{MSSM}} + \frac{\lambda_s \,
  m_{\textrm{SUSY}}}{M} \, (H_u \, H_d)^2 \; ,
\eeq
encoding the leading supersymmetric and $F$-term supersymmetry
breaking corrections to the Higgs sector that arise from a new
threshold at mass scale
$M$\,~\cite{Strumia:1999jm,Brignole:2003cm,Pospelov:2006jm,Antoniadis:2008es,Batra:2008rc}\,.
The corrections enter the spectrum and interactions through the
dimensionless parameters
\beq
\epsilon_{1} \equiv \frac{\lambda \,
  \mu^*}{M} \; , \qquad \epsilon_{2} \equiv \, - \, \frac{\lambda_s \,
  m_{\textrm{SUSY}}}{M} \; .
\eeq
For $M \sim$ few TeV, the BMSSM has interesting implications for
cosmology~\cite{Blum:2008ym,Cheung:2009qk,Bernal:2009hd,Berg:2009mq,Bernal:2009jc}
and for Higgs phenomenology~\cite{Carena:2009gx}.  The BMSSM
operators, which contribute at tree-level to the Higgs mass, alleviate
the tension associated with the stop sector and $\tan\beta$. Now, the
left-handed stop can also be relatively light, providing additional
bosonic degrees of freedom that strengthen the first-order phase
transition\footnote{To be clear, a first-order phase transition is
  induced radiatively through thermal effects, similar to the MSSM, as
  opposed to new tree-level interactions as in, {\it e.g.}, the
  NMSSM~\cite{NMSSM}.}.

In this work, we examine the BMSSM implications for CP violation and
the generation of the baryon asymmetry.  In Sec.~\ref{sec:CPv}, we
describe new CP-violating phases associated with the BMSSM operators.
In Sec.~\ref{sec:ewb}, we review relevant aspects of the phase
transition dynamics and show that these phases lead to new
CP-violating sources that generate charge density during the phase
transition. In this section we also compute the resulting BAU.  In
Sec.~\ref{sec:edms}, we discuss how searches for EDMs constrain CP
violation and baryogenesis in the BMSSM.  We conclude in
Sec.~\ref{sec:conclude}. The appendices contain details of the
CP-violating vacuum structure and radiative corrections to the Higgs
CP-violating phase.

\section{CP violation}
\label{sec:CPv}
In this section, we describe the BMSSM Lagrangian to leading order in
$M^{-1}$, emphasizing those aspects that are relevant for CP
violation, baryogenesis, and EDMs.  The new BMSSM phases (denoted
$\vartheta_{1,2}$) lead to (i) explicit CP violation in the
neutralino, chargino, and squark mass matrices, and (ii) CP-violating
mixing of the Higgs pseudoscalar $A_0$ with the two other neutral
Higgs scalars $h_0, H_0$.  We express our results in terms of physical
CP-violating phases that are invariant under phase redefinitions of
fields, summarized in Table~\ref{tab:phases}.

\begin{table}[t!]
\begin{tabular}{|c|c|c|c|c|}
\hline
\multicolumn{2}{|c}{MSSM phases} & \multicolumn{2}{|c|}{BMSSM phases}
& vev phase\\
\hline
$\phi_{i}$ &
$\phi_{f}$ & $\vartheta_1$ & $\vartheta_2$ & $\theta$\\
\hline
$ \; \arg(M_i \mu/b \; )$ & $ \; \arg(A_f \mu/b) \; $ & $\;
\arg(\epsilon_1/b) \; $ & $ \; \arg(\epsilon_2/b^2) \; $ &  $ \;
\arg(b \, H_uH_d) \; $ \\
\hline
\end{tabular}
\caption{\it\small The CP-violating phases in the BMSSM. Here
  $i=1,2,3$ labels gaugino mass parameters and $f$ labels the
  trilinear sfermion-Higgs coupling corresponding to a SM
  fermion $f$.
\label{tab:phases}}
\end{table}

Neglecting flavor mixing, the invariant phases of
Table~\ref{tab:phases} provide a complete basis for all of the
rephasing invariants in the BMSSM. Such basis is easily constructed by
noting that Higgs field rephasing is equivalent to global U(1)$_{\rm
  PQ}$ and U(1)$_{\rm R-PQ}$ transformations \cite{Dimopoulos:1995kn}.
The U(1)$_{\rm PQ}$ and U(1)$_{\rm R-PQ}$ are explicitly broken by the
dimensionful MSSM parameters appearing in Table~\ref{tab:phases}, as
well as by the BMSSM new effective couplings. By promoting the
parameters to spurions with well-defined transformation properties,
one can extract the rephasing invariants in terms of U(1)$_{\rm PQ}$
and U(1)$_{\rm R-PQ}$ conserving combinations.

We follow the notation of Ref.~\cite{Martin:1997ns} with respect to
the MSSM parameters.  In our numerical analysis we implement the
quantum corrections from the neutralino, chargino, scalar Higgs, and
squark sectors. Details are given in Appendix~\ref{app:QC}. CP
violation induced by these corrections in the Higgs sector is
suppressed for small values of the trilinear $A$ term and for moderate
values of $\mu$, which we adopt throughout our analysis. This allows
us to focus on the novel tree-level BMSSM effects.

First, we consider the tree-level Higgs potential
\begin{align}\label{eq:V0}
V_0 =& \left( m_{H_u}^2 + |\mu|^2 \right) \, \left|H_u\right|^2 +
\left( m_{H_d}^2 + |\mu|^2 \right) \, \left|H_d\right|^2 +
\frac{g^{\prime 2} + g^2}{8} \, \left( \left|H_u\right|^2 -
  \left|H_d\right|^2  \right)^2 + \frac{g^2}{2} \, \left| H_d^\dagger
  \, H_u \right|^2 \notag \\
& + \left( \frac{}{} b \, (H_u H_d) + 2 \, \epsilon_1 \, \left(
    \left|H_u\right|^2 + \left|H_d\right|^2  \right) (H_u H_d) +
  \epsilon_2 \, (H_u H_d)^2 + \; \textrm{h.c.} \, \right)  \; ,
\end{align}
with SU(2)$_L$ contractions defined as $(H_u H_d) \equiv H_u^+ H_d^-
\!- H_u^0 H_d^0$\,. At zero temperature, the Higgs vacuum expectation
values (vevs) are
\beq\label{eq:vev}
\left\langle H_u^0 \right\rangle \equiv v_u = s_\beta \, v \, e^{i \,
\theta_u} \; , \quad \left\langle H_d^0 \right\rangle \equiv v_d =
c_\beta \, v \, e^{i \, \theta_d}
\eeq
where $\tan\beta \equiv |v_u/v_d|$, $s_\beta\equiv\sin\beta$,
$c_\beta\equiv\cos\beta$, and $v \simeq 174$ GeV.  The relative phase
  $(\theta_u\!-\!\theta_d)$ is unphysical and can be set to zero by
  a gauge transformation. We define the Higgs
phase $\theta$ as
\beq
\theta_u + \theta_d \equiv \theta - \arg(b) \; .
\eeq
It is useful to factor out $\arg(b)$ explicitly, since $\theta$ is
rephasing invariant~\cite{Choi:2000wz}~\footnote{In
  Ref.~\cite{Ham:2009zn}, explicit CP violation in the BMSSM was also
  studied. However, the authors considered a scenario in which
  $\arg(b)=\theta_u + \theta_d=0$. Since $\theta$ is rephasing
  invariant, Ref.~\cite{Ham:2009zn} deals with a very specific
  physical model, and the results derived there do not apply in
  general.}. Next, we define rephasing invariant BMSSM parameters
\begin{subequations}
\label{eq:epsilondef}
\begin{align}
\epsilon_{1r} &\equiv |\epsilon_{1}| \, \cos(\vartheta_{1}+\theta) &
\epsilon_{1i} &\equiv |\epsilon_{1}| \, \sin(\vartheta_{1}+\theta) \\
\epsilon_{2r} &\equiv |\epsilon_{2}| \, \cos(\vartheta_{2}+2\theta)&
\epsilon_{2i} &\equiv |\epsilon_{2}| \, \sin(\vartheta_{2}+2\theta) \; .
\end{align}
\end{subequations}
In the phase convention $\theta_u+\theta_d = 0$, our definitions
reduce to the usual definitions $\epsilon_{1r} =
\textrm{Re}[\epsilon_1]$, {\it etc.}~\cite{Dine:2007xi}.

The masses and mixing angles of the physical Higgs bosons receive
tree-level corrections from the BMSSM operators. In our expressions to
follow, we work at tree-level and eliminate the set of parameters
$(m_{H_u}^2,m_{H_d}^2, |b|)$ in favor of $(v, \tan\beta, m_A)$.  We
parametrize the Higgs fields in the following way:
\beq\label{eq:Hdef}
H_u = e^{i \, \theta_u} \left( \begin{array}{c} H_u^+ \\
s_\beta \, v + \frac{h_u + i \, a_u}{\sqrt{2}} \end{array}\right) \; ,
\quad H_d = e^{i \, \theta_d} \left( \begin{array}{c}
    c_\beta \, v + \frac{h_d + i \, a_d}{\sqrt{2}} \\
    H_d^- \end{array} \right) \; .
\eeq
In the limit $\epsilon_{1i} = \epsilon_{2i} = 0$, one can separately
diagonalize the CP-even and odd Higgs states, as in the
MSSM~\cite{Martin:1997ns}. The eigenstates are
\beq
\left( \begin{array}{c} h_0 \\ H_0 \end{array} \right) = \left(
  \begin{array}{cc} \cos\alpha & - \sin\alpha \\  \sin\alpha &
    \cos\alpha \end{array} \right) \left( \begin{array}{c} h_u \\ h_d
  \end{array} \right) \; , \qquad
\left( \begin{array}{c} G_0 \\ A_0 \end{array} \right) = \left(
  \begin{array}{cc} \sin\beta & -\cos\beta \\ \cos\beta & \sin\beta
  \end{array} \right) \left( \begin{array}{c} a_u \\ a_d \end{array}
\right) \; ,
\eeq
with Higgs mixing angle $\alpha$ given by
\beq
\cos 2\alpha = - \frac{ m_A^2 - m_Z^2
+ 4 \epsilon_{2r} v^2}{m_{H}^2- m_h^2} \;  \cos 2\beta \; , \qquad
\sin 2\alpha = - \frac{ (m_A^2 + m_Z^2) \sin 2\beta
- 8 \epsilon_{1r} v^2 }{m_{H}^2 - m_h^2}\; .
\eeq
The mass eigenvalues also receive tree-level
contributions proportional to $\epsilon_{1,2r}$~\cite{Dine:2007xi}.
In particular, the lightest Higgs boson receives a correction
\beq\label{eq:higgsDST}
\delta_\epsilon m^2_{h}  = 2 \, v^2 \, \left( \epsilon_{2r} - 2 \,
  \epsilon_{1r} s_{2\beta} - \frac{ 2 \epsilon_{1r} (m_A^2 + m_Z^2)
    s_{ 2\beta} + \epsilon_{2r} (m_A^2 - m_Z^2) c^2_{ 2\beta}
}{\sqrt{ (m_A^2 - m_Z^2)^2 + 4 m_A^2 m_Z^2 s^2_{ 2\beta} } } \right)
\; .
\eeq
This contribution can increase the tree-level Higgs mass above the LEP
bound, without the need for radiative corrections~\cite{Blum:2009na}.
It is important to recall that the LEP bound on the lightest neutral
Higgs boson mass is drastically changed in the presence of CP
violation \cite{Schael:2006cr}.  In fact, there are allowed regions
even for very small values of the Higgs boson mass.  In principle this
implies that a much larger region in parameter space can accommodate a
strong first order phase transition and the right-handed stop can be
heavier than the top quark. However, as we will see below, there are
significant constraints on the amount of CP violation.

CP violation enters the Higgs sector at tree-level when
$\epsilon_{1,2i} \ne 0$, leading to mixing between CP eigenstates.  In
the $(h_0, H_0, A_0)$ basis, the Higgs mass matrix is
\beq\label{eq:m2Hpartdiag}
M_{H_0}^2 \; = \;  \left(\begin{array}{ccc} m_{h}^2  & 0 & m^2_{hA} \\
    0 & m_{H}^2 & m^2_{HA} \\ m^2_{hA} & m^2_{HA} & m_A^2 \end{array} \right)\;.
\eeq
The remaining CP-odd state $G_0$ is eaten by the $Z$ boson.
The parameters $m^2_{hA}$ and $m^2_{HA}$, which govern the mixing
between CP-even and odd states, are given by
\beq
m^2_{hA} &=& 4 v^2 \epsilon_{1i} \, \sin(\beta - \alpha) - 2 v^2
\epsilon_{2i} \, \cos(\alpha+\beta)
\;\approx \;-2 v^2 \,\left(\epsilon_{2i} \, s_{2\beta} - 2 \epsilon_{1i}\right)
\; , \\
m^2_{HA} &=& 4 v^2 \epsilon_{1i} \, \cos(\beta - \alpha) - 2 v^2
\epsilon_{2i} \, \sin(\alpha+\beta)  \;
\approx \; 2 v^2 \epsilon_{2i} \, c_{ 2 \beta} \; ,
\eeq
where the approximations follow in the limit of moderate $\tan\beta$
and $m_A^2 \gg m_Z^2$, such that $\alpha \approx \beta - \pi/2$.  To
$\mathcal{O}(\epsilon_{1,2i})$, the eigenvalues are unchanged from the
CP-conserving case.  Note that $m_A$ is now the mass of the
``mostly-pseudoscalar'' eigenstate, not the mass of $A_0$.  To avoid
this confusion, we will express physical quantities in terms of the
charged Higgs boson mass, using the relation
\beq\label{eq:mAmHctree}
m_{H_\pm}^2 = m_A^2 + m_W^2 + 2 \, \epsilon_{2r} \, v^2 \; .
\eeq

The Higgs phase $\theta$, determined by the minimization condition
$\partial V_0/\partial\theta = 0$, is given by
\beq\label{eq:imagmin1}
\tan\theta \; = \; \frac{ 2\,v^2\,\left( \epsilon_{2i} \, s_{
      2\beta}\,-\, 2 \, \epsilon_{1i}\right)}{s_{ 2\beta} \, (
  m_{H_\pm}^2 - m_W^2 ) \,+\,2\,v^2\,\left( \epsilon_{2r} \, s_{
      2\beta}\,-\, 2 \, \epsilon_{1r}\right) } \; .
\eeq
In the small $\tan\beta$ regime (such that $\cot\beta \gg
|\epsilon_1|v^2/m_A^2$), one can treat $\theta$ perturbatively since
it is $\mathcal{O}(\epsilon_{1,2})$\,.  However, in the large
$\tan\beta$ regime (such that $\cot\beta \lesssim
|\epsilon_1|v^2/m_A^2$\,)\,, one can have $\theta=\mathcal{O}(1)$\,.
In this regime, the Higgs potential can develop more than one minimum
in the $\theta$ direction; we discuss this possibility in
Appendix~\ref{app:thtree}.  In practice, constraints on $b\to
s\,\gamma$~\cite{Alam:1994aw} imply that the mass of the charged Higgs
cannot be too light ($m_{H_\pm}\gtrsim300$\,GeV) unless the charged
Higgs contribution to $b\to s\,\gamma$ interferes destructively with
some other process~\footnote{Some amount of interference is in fact
  expected, considering the light stops and charginos of our
  baryogenesis scenario, weakening the bound on $m_{H_\pm}$.}.
Restricting ourselves to $m_{H_\pm}>200$\,GeV and $\tan\beta < 10$ is
sufficient to avoid additional phase minima.

CP violation from the complex BMSSM parameters $\epsilon_{1,2}$ also
enters the SUSY mass matrices, potentially impacting both baryogenesis
and EDMs.  For example, the top squark mass matrix, in the
$(\widetilde t_L^*, \, \widetilde t_R^*)$ basis, is
\beq\label{eq:mstop}
{\bf m^2_{\widetilde t} } = \left(
\begin{array}{cc} m_{Q_3}^2 + y_t^2 s_\beta^2 v^2 + \Delta_{\widetilde
 u_L} & y_t (A_t^* v_u^* - \mu v_d + 2 \epsilon_1 v_u v_d^{2}/\mu^* ) \\
  y_t (A_t v_u - \mu^* v_d^* + 2 \epsilon^*_1 v_u^* v_d^{*2}/\mu ) &
  m_{\bar{u}_3}^2 + y_t^2 s_\beta^2 v^2 + \Delta_{\widetilde u_R}
  \end{array} \right)
\eeq
with D-term contributions $\Delta_\phi = \left( T_{3\phi} - Q_\phi \,
  \sin^2 \theta_W \right) c_{ 2\beta} \, m_Z^2$.  We define the
stop mixing parameter $X_t = |[{\bf m^2_{\widetilde t}} ]_{12} |$ as
the magnitude of the off-diagonal entry in Eq.~\eqref{eq:mstop}.  The
chargino mass matrix, in the $(\widetilde W^+, \, \widetilde H_u^+, \,
\widetilde W^-, \, \widetilde H_d^- )$ basis, is
\beq
{\mathbf M}_{\widetilde C} = \left( \begin{array}{cc} 0 & \mathbf{X}^T
    \\ \mathbf{X} & 0 \end{array} \right) \;, \qquad \mathbf X =
\left( \begin{array}{cc} M_2 & g \, v_u^* \\ g \, v_d^* & \; \mu
    - 2 \epsilon_1 v_d v_u /\mu^* \end{array} \right) \; .
\eeq
The neutralino mass matrix, in the $( \widetilde B,
\, \widetilde W^0, \, \widetilde H_d^0, \, \widetilde H_u^0 )$ basis,
is
\beq
{\bf M}_{\widetilde N} = \left(\begin{array}{cccc}
M_1 & 0 & - g^\prime v_d^* /\sqrt{2} &  g^\prime v_u^* /\sqrt{2} \\
0 & M_2 & g v_d^*/\sqrt{2} & - g v_u^*/\sqrt{2} \\
- g^\prime v_d^* /\sqrt{2} & g v_d^*/\sqrt{2} & 2 \epsilon_1 v_u^2
/\mu^* & - \mu + 4 \epsilon_1 v_u v_d /\mu^* \\
g^\prime v_u^*/\sqrt{2} & - g v_u^*/\sqrt{2} & - \mu + 4 \epsilon_1
v_u v_d /\mu^* & 2 \epsilon_1 v_d^2 /\mu^*
\end{array} \right) \; .\label{eq:neutralinomass}
\eeq
In each of these mass matrices, the BMSSM parameters lead to new
sources of CP violation through both the explicit factors of
$\epsilon_1$ and the complex phase of the Higgs vevs, depending on
$\epsilon_{1}$ and $\epsilon_{2}$\,. Physical CP violation observables
depend on these phases only through the invariant combinations listed
in Table~\ref{tab:phases}.

To summarize, CP violation from the BMSSM manifests in the following
ways:
\begin{itemize}
\item Mixing arises between CP-even and CP-odd neutral Higgs
  eigenstates, proportional to $m^2_{hA}$ and $m^2_{HA}$.
\item The Higgs phase $\theta$ enters through the Higgs vevs in the
  SUSY mass matrices.
\item The parameter $\epsilon_{1}$ appears explicitly in the SUSY mass
  matrices.
\end{itemize}
In Sec.~\ref{sec:ewb}, we show that the BAU induced through BMSSM
phases is approximately proportional to $\theta$.  In
Sec.~\ref{sec:edms}, we find that the dominant contributions to EDMs
arise through either $\theta$ or $h_0$-$A_0$ mixing.  Since $\theta$
and $m^2_{hA}$ are proportional to the same linear combination of
$\epsilon_{1,2i}$, EDM constraints will provide direct bounds on the
EWBG mechanism in the BMSSM.

\section{Electroweak Baryogenesis}
\label{sec:ewb}
In this section, we describe how electroweak baryogenesis is realized
in the BMSSM.  First, we study the nature of the phase transition and
the properties of the expanding bubbles relevant for the BAU
computation.  Second, we identify novel sources of CP violation in the
BMSSM and compute the resulting BAU.  The new sources are induced by a
variation of the Higgs phase $\theta$ across the bubble wall.
Therefore, we devote special attention to the computation of the
temperature and space-time dependence of $\theta$. Similar effects
arise also in the MSSM at the quantum
level~\cite{CPvbubmssm,Huber:1999sa}. In the BMSSM they arise
classically~\cite{Blum:2008ym}, and can be quantitatively much more
significant.

\subsection{Phase Transition and Bubble Properties}\label{ssec:wall}

\subsubsection{The critical vev and temperature}\label{sssec:vctc}
The EWBG mechanism requires a ``strong'' first-order phase transition
to avoid sphaleron erasure of the BAU within the broken phase.  This
condition is satisfied if~\cite{Shaposhnikov:1986jp}
\beq\label{eq:vcTc1}
\frac{\sqrt{2} \, v_c}{T_c}>1\;,
\eeq
where $T_c$ is the critical temperature (defined here as the
temperature of free-energy degeneracy between the broken and symmetric
phases) and $v_c \equiv v(T_c)$ is the Higgs vev in the broken phase
at $T_c$, in the normalization of \eq{eq:Hdef}.
Since a first-order phase transition in the BMSSM arises through
radiative corrections involving stops (as in the
MSSM~\cite{mssmstops,Quiros:2000wk}), Eq.~\eqref{eq:vcTc1} provides
important constraints on the parameters of the stop sector.  In
addition, the size of $v_c$ itself is important for the BAU
computation; as we show below, the BAU scales as $v_c^4$.

\begin{figure}[!t]
\begin{center}
\mbox{\hspace*{0cm}\epsfig{file=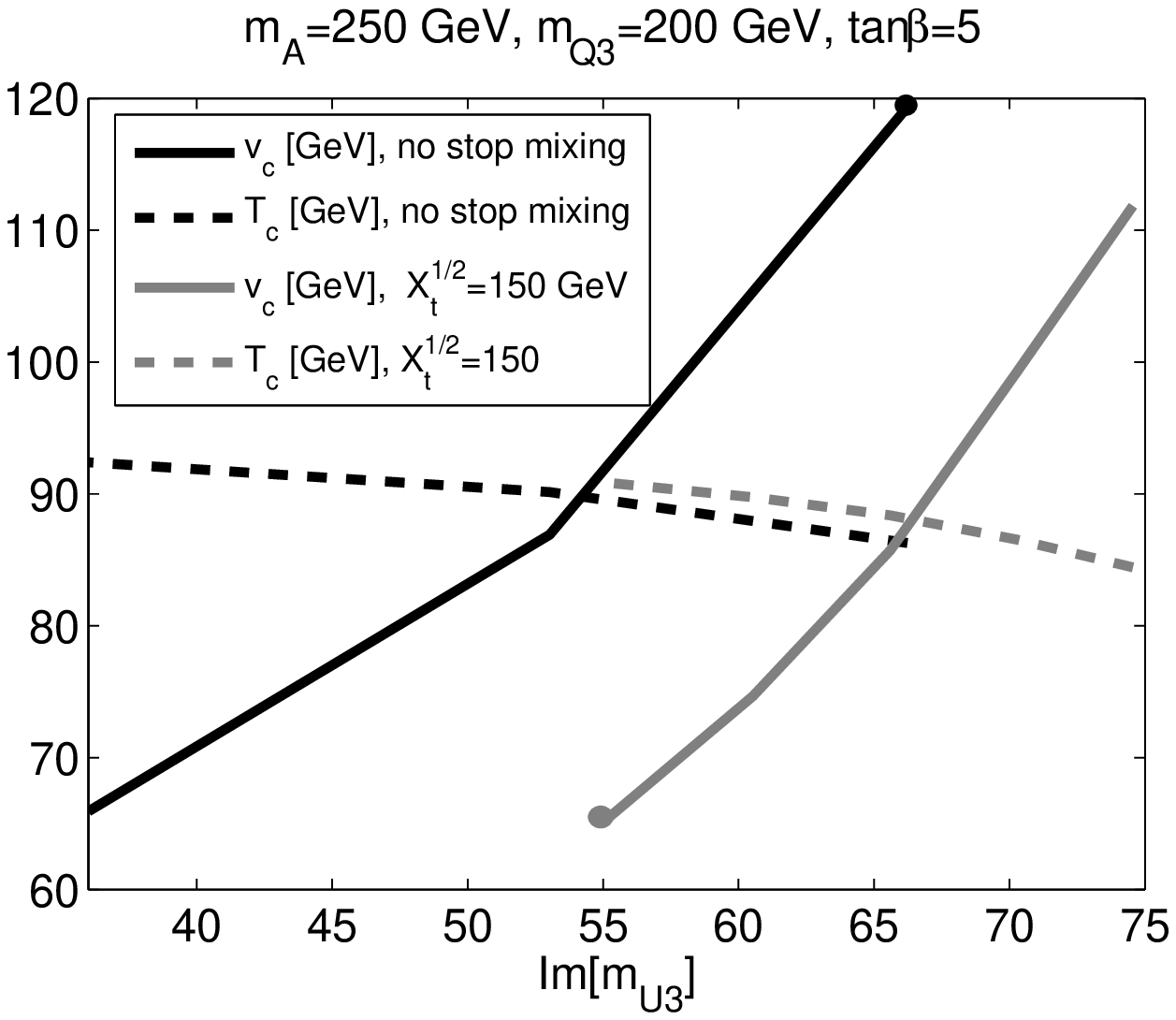,height=6cm}}
\mbox{\hspace*{0cm}\epsfig{file=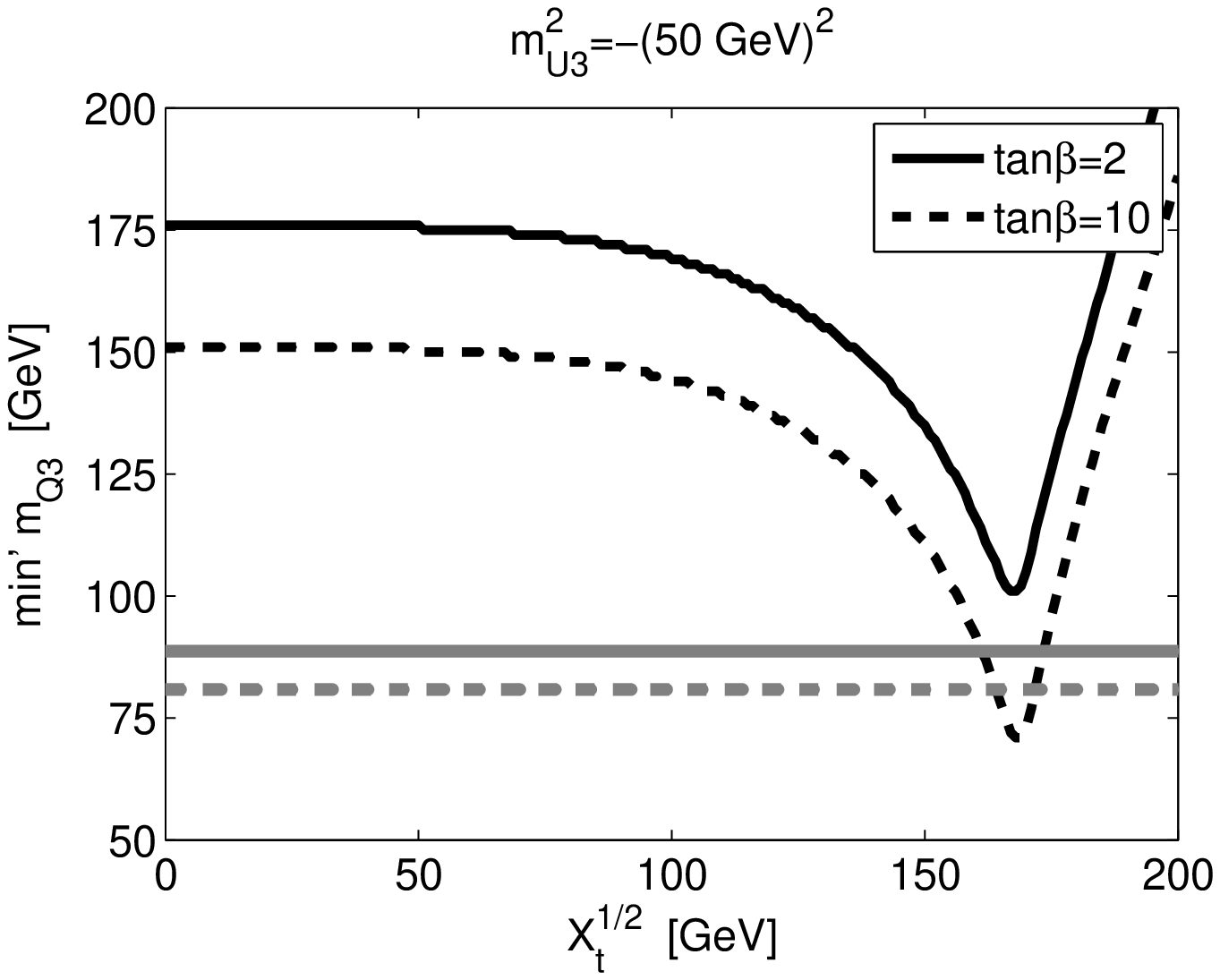,height=6cm}}
\end{center}
\caption{\it Left: The critical vev $v_c$ (solid curve) and
  temperature $T_c$ (dashed) vs.~the right-handed stop soft mass $m_{U_3}$
  (imaginary, since $m_{U_3}^2 < 0$), for two values of the mixing parameter
  $X_t$\,. Right, black: The minimum value of $m_{Q_3}$,
  consistent with the electroweak oblique parameter $T<0.2$, as a
  function of the stop mixing $\sqrt{X_t}$, for two values of
  $\tan\beta$. Gray: Model independent, direct limit from sbottom search.}
\label{fig:vcTc}
\end{figure}

We compute $v_c$ and $T_c$ using the two-loop finite-temperature
effective potential of Ref.~\cite{Bernal:2009hd}, provisionally
neglecting the effect of CP violation.  In Fig.~\ref{fig:vcTc} (left
panel), we show how $T_c$ and $v_c$ depend on the stop parameters
[{\it c.f.}~Eq.~\eqref{eq:mstop}].  We consider two cases: mixing
($X_t=\,(150\,{\rm GeV})^2$) and no mixing ($X_t=0$), while varying
$m_{U_3}^2$ (assuming $m_{U_3}^2 < 0$) and fixing $m_A=250$\,GeV,
$\tan\beta=5$, $m_h=114$\,GeV, and $m_{Q_3}=200$\,GeV. At a given
value of $m^2_{U_3}$, stop mixing suppresses $v_c/T_c$, thereby
weakening the phase transition. The filled circle at the lower edge of
the $X_t>0$ line (gray) corresponds to the maximal value of
$m_{U_3}^2$ where \eq{eq:vcTc1} is fulfilled. On the other hand,
decreasing $m_{U_3}^2$ increases $v_c/T_c$, strengthening the phase
transition; eventually, however, this leads to an undesirable
tachyonic stop, denoted by the filled circle at the upper edge of the
$X_t=0$ line (black). These effects are well known in the MSSM: stop
mixing effectively screens the one-loop cubic correction to the
effective potential and increases the value of the Higgs mass, while
such screening can be compensated by a sufficiently negative
$m_{U_3}^2$.

Next, we consider how the strength of the phase transition depends on
$m_{Q_3}$.  We vary $m_{Q_3}$ while keeping the neutral Higgs mass
fixed by simultaneously adjusting $\epsilon_{1,2r}$.  For $m_{Q_3} >
200$ GeV, $\, v_c$ and $T_c$ are only weakly dependent on $m_{Q_3}$.
For $m_{Q_3} < 200$ GeV, the phase transition is strengthened as one
decreases $m_{Q_3}$, allowing for Eq.~\eqref{eq:vcTc1} to be
consistent with greater values of
$m_{U_3}^2$~\cite{Blum:2008ym,Bernal:2009hd}.  However, experimental
constraints provide a lower bound on $m_{Q_3}$, shown in
Fig.~\ref{fig:vcTc} (right panel).  The collider search bound on
bottom squarks ($m_{\widetilde b} > 89$ GeV~\cite{Amsler:2008zzb})
implies that $m_{Q_3} \gtrsim 80$ GeV~\cite{Bernal:2009hd}, shown in
gray~\footnote{For a neutralino LSP with mass $m_{\chi_1^0} \lesssim
  90$ GeV, the bound is significantly stronger: $m_{\widetilde b}
  \gtrsim 250$ GeV~\cite{D0note}.}.  The precision constraint on the
$\rho$ parameter, which is sensitive to the
$\mathcal{O}(m_t^2/m_{Q_3}^2)$ relative mass splitting in the doublet
$(\tilde t_L,\,\tilde b_L)$, provides a stronger
bound~\cite{Drees:1990dx,Amsler:2008zzb}.  We find that
$m_{Q_3}\gtrsim160$\,GeV is required at the 95\% CL, unless the stops
are substantially mixed; at the 90\% CL one must keep
$m_{Q_3}\gtrsim200$\,GeV.

\subsubsection{The bubble profiles}\label{sssec:bub}
We now turn to the properties of the bubbles that nucleate and expand
during the phase transition~\cite{Moreno:1998bq,Huber:2001xf}.  These
bubbles are characterized by spacetime-dependent background Higgs
fields, described by $(v, \, \beta, \, \theta)$.  In the MSSM, the
dominant sources driving EWBG are proportional to the small parameter
$\Delta \beta \lesssim 10^{-2}$ associated with the spacetime
variation of $\beta$~\cite{Moreno:1998bq}.  Here, we neglect variation
in $\beta$ ($\Delta\beta = 0$) in order to focus on the leading BMSSM
effects proportional to the analogous parameter $\Delta\theta$
associated with variation of $\theta$ across the bubble wall.

The bubble profiles for $v$ and $\theta$ are determined by equations
of motion
\begin{subequations}\beq
\frac{\partial V_T}{\partial v}&=&2\,\frac{\partial^2 v(z)}{\partial
  z^2}\label{eq:vprofile}\\
\frac{\partial V_T}{\partial
  \theta}&=&\frac{s^2_{2\beta}}{2}\,\partial_z\left(v^2(z)\,\partial_z\,
 \theta(z)\right)
  \label{eq:thprofile}\;,
\eeq\end{subequations}
assuming a vanishing $Z$
background~\cite{Comelli:1995em,Huber:1999sa}.  Here, $V_T$ is the
finite temperature effective potential and the coordinate $z$ is the
distance from the wall in its rest frame (we assume a planar bubble).
The boundary conditions are such that $z=+\infty \, (-\infty)$
corresponds to the broken (unbroken) minimum of the potential.

Rather than solving Eq.~\eqref{eq:vprofile}, we assume a kink ansatz
for the Higgs vev:
\beq
v(z) &=& \frac{v(T)}{2} \left[ 1 + \textrm{tanh}\left( \frac{2\,
 \, z}{L_w}  \right) \, \right] \label{eq:vz}\;.
\eeq
To avoid complication, we set $v(T)=v_c$. This potentially
underestimates the BAU: since the time of bubble growth necessarily
corresponds to $T<T_c$, one in general expects $v(T)>v_c$, resulting
with more effective sources. The approximation is justified if,
following the onset of the phase transition, the universe is reheated
back near the critical
temperature~\cite{KurkiSuonio:1996rk,Megevand:2000da}.

The wall width $L_w$ is defined to match the kink ansatz of \eq{eq:vz}
onto the bubble profile,
\beq\label{eq:Lw}
L_w=\int_{v_{\rm min}}^{v_{\rm max}}\frac{d\phi}{\sqrt{V_T(\phi)}}\;,
\eeq
where $v_{\rm min}=0.1\,v_c$ and $v_{\rm max}=0.9\,v_c$ designate the
field on either side of the wall.  Neglecting CP violation in
Eq.~\eqref{eq:Lw}, we find values in the range $L_w=(15-40)/T_c$.

Next, we obtain the profile of the vev phase $\theta(z)$.  In order to
clearly illustrate the essential dynamics, we make two
simplifications: (i) we consider the $\theta \ll 1$ regime, satisfied
when $\cot\beta \gg |\epsilon_1|v^2/m_A^2$, and (ii) we neglect all
radiative corrections to the phase-dependent part of $V_T$.  Under
these assumptions, Eq.~\eqref{eq:thprofile} becomes
\beq\label{eq:theom}
\partial_z\left( \frac{}{} v^2(z)\:\partial_z\,\theta(z) \, \right) \;
\approx \; v^2(z)\,
(m_{H_\pm}^2- m_W^2) \,\left[\,
  \theta(z)-\frac{v^2(z)}{v_0^2}\,\theta_0 \, \right]\;.
\eeq
In the present section, the zero temperature Higgs phase and vev are
denoted $\theta_0$ and $v_0$ for clarity.  Eq.~\eqref{eq:theom} can be
cast in the dimensionless form
\beq\label{eq:feom}
\partial_r\left(g \, \partial_rf\right)\; = \; \tau \, g\, (f-g)\;,
\eeq
where
\beq
f(z) \equiv\frac{v^2_c\,\theta(z)}{v_0^2 \, \theta_0} \:, \; \;
r\equiv \frac{2
  z}{L_w}\,,\;\;\tau\equiv\frac{L_w^2(m_{H_\pm}^2-m_W^2)}{4}\,,\;\;
g(r)\equiv\left(\frac{1+\tanh r}{2}\right)^2.
\eeq
For $m_{H_\pm} > 200$ GeV, we have $\tau > 100$.  Thus, the solution
to Eq.~\eqref{eq:feom} is governed by the potential energy term on the
RH side, such that $f(r) \approx g(r) + \mathcal{O}(\tau^{-1})$.  This
conclusion is borne out by numerical evaluations for $\tau > 30$,
which we perform using the method of Ref.~\cite{John:1998ip}.

In summary, the Higgs phase profile is given by
\beq\label{eq:thetaz}
\theta(z) \; \approx \; \frac{\Delta\theta}{4} \, \left[ 1 +
  \textrm{tanh}\left( \frac{2\,
 \, z}{L_w}  \right) \, \right]^2
\eeq
where $\Delta\theta \approx \theta_0 v_c^2/v_0^2$.  We note that
$\theta(z)$ is proportional to the square of the kink in
Eq.~\eqref{eq:vz}, not linear.  Furthermore, since EDMs are directly
sensitive to the value of $\theta_0$, this sensitivity translates into
a direct constraint on the phase variation across the bubble wall, and
hence on baryogenesis.

The preceeding analysis can be generalized away from the $\theta \ll
1$ regime, necessary when $\tan\beta \gtrsim 10$.  Assuming that the
profile of $\theta$ is again dominated by potential energy (such that
$\partial V_T/\partial\theta \approx 0$), we find the following
approximate solution for $\theta(z)$:
\beq
\tan\left(\theta(z) - \theta_0\right) \; \approx \; \frac{ 2 \,
  (s_{2\beta} \, \epsilon_{2i} - 2\, \epsilon_{1i})\, (v(z)^2 - v^2_0)
}{s_{2\beta}\, (m_{H_\pm}^2 - m_W^2) + 4\, \epsilon_{1r}\, (v(z)^2 -
  v_0^2) }  \; .
\eeq
This solution gives the leading behavior of $\theta(z)$ in all
$\tan\beta$ regimes: (i) it reduces to Eq.~\eqref{eq:thetaz} when
$\cot\beta \gg |\epsilon_1|v^2/m_A^2$ limit, and (ii) it is valid to
leading order in $\cot\beta$ when $\cot\beta \lesssim
|\epsilon_1|v^2/m_A^2$.

In Appendix~\ref{app:QC}, we study the impact of radiative corrections
on the Higgs phase at zero and finite temperature.  In particular, we
find that $\theta(z)$ is shifted by an overall constant, while
$\Delta\theta$ remains approximately unchanged.

\subsubsection{Wall velocity}
The bubble wall velocity is an important parameter in the EWBG
computation.  A recent study found that $v_w\sim 0.4$ in the
MSSM~\cite{Megevand:2009gh}, significantly larger than previous
estimates of $v_w\sim\;0.01\!-\!0.1$~\cite{bubvelocity}.  Therefore it
is worthwhile examining how the BAU depends on $v_w$.

The optimal wall velocity for EWBG arises as a competition between two
Sakharov conditions.  The generation of baryon number ($n_B$) is
fueled by chiral charge diffusing ahead of the advancing bubble wall,
characterized by an effective diffusion constant $\bar D$ and a diffusion
time $\tau_{\textrm{diff}} = \bar{D}/v_w^2$~\cite{Chung:2009cb}.  If
electroweak sphalerons are in equilibrium, with rate
$\Gamma_{\textrm{ws}} \, \gg \, \tau_{\textrm{diff}}^{-1}$, $n_B$ is
suppressed, as per the third Sakharov condition.  On the other hand,
if $\Gamma_{\textrm{ws}} \, \ll \, \tau_{\textrm{diff}}^{-1}$, then
$n_B$ is also suppressed, since few baryon number violating processes
occur.  Therefore, the maximum baryon number production occurs when
$\Gamma_{\textrm{ws}} \sim \tau_{\textrm{diff}}^{-1}$, corresponding
to a velocity $v_w \sim \sqrt{ \bar{D} \Gamma_{\textrm{ws}} } \sim
(\textrm{few})\times 10^{-2}$~\cite{Megevand:2000da}.  In our
numerical computation, described below, we indeed find that $n_B$ is
maximized for $v_w = 0.03$.

If we consider the range $0.01 < v_w < 0.4$, we find $n_B$ varies by a
factor of $4\!-\!5$, with the minimum $n_B$ for $v_w = 0.4$.  For the
sake of definiteness, we fix $v_w = 0.1$, which is approximately the
central value for $n_B$.  We expect that $v_w$ in the BMSSM can be
approximated by the MSSM case.  Potentially, the presence of the light
LH stop leads to an additional contribution to the frictional force
determining $v_w$.  However, we expect this to be a minor effect since
$\widetilde t_L$, which cannot be too light, is somewhat Boltzmann
suppressed and acquires only a fraction of its mass via the Higgs
mechanism.

\subsection{Baryon Asymmetry Computation}\label{sec:BG}
The computation of the BAU involves a system of coupled Boltzmann
equations of the form
\beq
\partial_t \, n_a - D_a \, \nabla^{\,2} \, n_a = \sum_{b} \,
\Gamma_{ab} \, n_b + S^\CPv_a \;.
\eeq
Here $n_a$ is the charge density for species $a$.  The CP-violating
source $S^\CPv_a$, non-zero only within the moving bubble wall, leads
to the generation of non-zero $n_a$.  The diffusion constant $D_a$
describes how efficiently $n_a$ is transported ahead of the wall into
the unbroken phase where sphalerons are active.  The interaction
coefficients $\Gamma_{ab}$ correspond to (i) inelastic processes that
convert charge from one species to another, and (ii) relaxation
processes that wash out charge within the broken phase.  Although
BMSSM contributions modify $\Gamma_{ab}$ at
$\mathcal{O}(\epsilon_{1,2})$, it is safe to neglect these
corrections.  Previous studies have shown that the solutions to the
Boltzmann equations are insensitive to sub-$\mathcal{O}(1)$ variations
in the interaction coefficients~\cite{Cirigliano:2006wh, Chung:2009cb,
Chung:2009qs}.  We refer the reader to Ref.~\cite{Cirigliano:2006wh,Chung:2009qs},
which we follow here, for details concerning the setup and derivation
of the Boltzmann equations in the MSSM.

\subsubsection{CP-violating sources}
The novelty of BMSSM baryogenesis appears in the CP-violating sources.
We compute these sources following the ``vev-insertion'' approach of
Refs.~\cite{Riotto:1998zb, Lee:2004we}. More sophisticated treatments,
going beyond the vev-insertion approximation, exist in the
literature~\cite{Prokopec, Carena:2002ss}.  However, there remains
some controversy, and this is an area of active
investigation~\cite{newsourcework}.  Therefore, we opt for the
simplest framework (vev-insertion) by which we may present our new
BMSSM sources.

The higgsino CP-violating source, which drives EWBG in the MSSM,
receives important BMSSM contributions.  We have
\begin{align}
S^\CPv_{\widetilde H}(z) = & \frac{3g^2 K_{2}(z)}{2 \pi^2} \,
\int^\infty_0 \!\! \frac{k^2 \, dk}{\omega_{\widetilde W}\,
  \omega_{\widetilde H} } \; \textrm{Im}\left[ \, \frac{
    n_F(\mathcal{E}_{\widetilde W}) - n_F(\mathcal{E}_{\widetilde
      H}^*) }{ ( \mathcal{E}_{\widetilde W} -  \mathcal{E}_{\widetilde
      H}^* )^2 } + \frac{ 1- n_F(\mathcal{E}_{\widetilde W}) -
    n_F(\mathcal{E}_{\widetilde H}^*) }{ ( \mathcal{E}_{\widetilde W}
    +  \mathcal{E}_{\widetilde H} )^2 } \, \right]
\label{eq:Higgsinosource}\\
& + \frac{g^{\prime 2} K_{1}(z)}{2 \pi^2} \, \int^\infty_0 \!\!
\frac{k^2 \, dk}{\omega_{\widetilde B} \, \omega_{\widetilde H} } \;
\textrm{Im}\left[ \, \frac{ n_F(\mathcal{E}_{\widetilde B}) -
n_F(\mathcal{E}_{\widetilde H}^*) }{ ( \mathcal{E}_{\widetilde B} -
\mathcal{E}_{\widetilde H}^* )^2 } + \frac{ 1-
n_F(\mathcal{E}_{\widetilde B}) - n_F(\mathcal{E}_{\widetilde H}) }{ (
\mathcal{E}_{\widetilde B} + \mathcal{E}_{\widetilde H} )^2 } \,
\right] \notag \; .
\end{align}
The first and second terms correspond to the sources induced through
higgsino-wino and higgsino-bino mixing, respectively. The important
BMSSM effects enter into the prefactors
\beq\label{eq:bmssmKi}
K_i(z) = \left|M_i \, \mu\right| \, v^2(z) \, \left[
  \, \sin\left(\phi_i + \theta(z) \right)\, \dot\beta(z) +
  \frac{s_{4\beta}}{4} \, \cos\left(\phi_i + \theta(z)  \right)\,
  \dot\theta(z) \, \right] \; ,\no\\
\eeq
where $M_{1,2}$ are the gaugino mass parameters. In the MSSM we
have $\theta = \dot\theta = 0$, so that the sources are driven by
the gaugino phases $\phi_i$ defined in Table~\ref{tab:phases}.
However, in the BMSSM, contributions arise from $\epsilon_{1,2i}$,
entering through $\theta$.  Futhermore, the second term in
Eq.~\eqref{eq:bmssmKi}, which is unique to the BMSSM, is not
suppressed by $\Delta\beta \lesssim 10^{-2}$. The momentum integrals
in Eq.~\eqref{eq:Higgsinosource} are identical to the MSSM case,
discussed in Ref.~\cite{Lee:2004we}; roughly speaking, they are
maximized ``on-resonance'' (when $|M_i| \sim |\mu|$) and are highly
suppressed far off-resonance.

In the MSSM, the CP-violating sources for third generation squarks
cannot drive EWBG. The Higgs mass bound requires that $\widetilde t_L$
and $\widetilde b_L$ are heavy, and thereby Boltzmann suppressed in
the electroweak plasma.  The BMSSM opens the door for squark-driven
baryogenesis, since the stops and sbottoms can be relatively light.
The CP-violating sources for stops (${\widetilde q}={\widetilde t}$)
and sbottoms (${\widetilde q}={\widetilde b}$) are
\beq
S^\CPv_{\widetilde q_R} &=& \; - \; S^\CPv_{\widetilde q_L} \\
\notag &=& \frac{3 y_q^2 K_{\widetilde q}(z)}{2 \pi^2} \,
\int^\infty_0 \!\!  \frac{k^2 \, dk}{\omega_{\widetilde q_L}\,
  \omega_{\widetilde q_R} } \; \textrm{Im}\left[ \, \frac{
    n_B(\mathcal{E}_{\widetilde q_R}^*) - n_B(\mathcal{E}_{\widetilde
      q_L}) }{ ( \mathcal{E}_{\widetilde q_L} -
    \mathcal{E}_{\widetilde q_R}^* )^2 } + \frac{ 1 +
    n_B(\mathcal{E}_{\widetilde q_R}) + n_B(\mathcal{E}_{\widetilde
      q_L}) }{ ( \mathcal{E}_{\widetilde q_L} +
    \mathcal{E}_{\widetilde q_R} )^2 } \, \right] \; ,
\eeq
with prefactor
\beq\label{eq:bmssmKqtilde}
K_{\widetilde q}(z) &=& \left|A_q \, \mu\right| \, v^2(z) \,
\dot\beta(z)\,\sin\left(\phi_q+\theta(z)\right) \,\no\\
 &+& \frac{v^2(z)}{4}\, \left( \, s_{4\beta} \: \left|A_q \,
 \mu\right| \,\cos\left(\phi_q+\theta(z)\right) + s^2_{
2\beta} \, \left( |\mu|^2 - |A_t|^2 \right) \,
\right) \, \dot\theta(z) \; .
\eeq
In addition to the squark phases $\phi_q$ appearing in
Table~\ref{tab:phases}, the CP-violating sources include contributions
from $\epsilon_{1,2i}$ that enter through $\theta$.

In the BMSSM, there are CP-violating sources for the third generation
quarks, top ($q=t$) and bottom ($q=b$), due to the Higgs phase
$\theta(z)$, that does not arise in the MSSM:
\beq
S^\CPv_{q_R} &=& \; - \; S^\CPv_{q_L} \;
= \frac{3 y_q^2 K_{q}(z)}{2 \pi^2} \, \int^\infty_0 \!\! dk \, k^2 \notag \\
&\;& \; \; \times \; \textrm{Im}\!\left[ \, Z_{q_L}^{p} Z_{q_R}^{h}
  \frac{ n_F(\mathcal{E}^{h*}_{q_R}) - n_B(\mathcal{E}^p_{q_L}) }{ (
    \mathcal{E}_{q_L}^p - \mathcal{E}^{h*}_{q_R} )^2 } + Z_{q_L}^{p}
  Z_{q_R}^{p} \frac{ 1 + n_B(\mathcal{E}^p_{ q_R}) +
    n_B(\mathcal{E}^p_{q_L}) }{ ( \mathcal{E}_{q_L}^p +
    \mathcal{E}_{q_R}^p )^2 } \, + \; ( p \leftrightarrow h) \;
\right] \; . \notag
\eeq
where
\beq
K_q(z) = - \, v^2(z) \, s^2_{2\beta} \, \dot\theta(z) \; .
\eeq
Again, we refer to Ref.~\cite{Lee:2004we} for the notation of
quantities within the momentum integral.  We note that because the
quark thermal masses and widths are approximately equal for $q_L$ and
$q_R$ (dominated by common QCD effects), the momentum integral is
suppressed. This situation may be an artifact of the vev-insertion
approach.  Similar quark CP-violating sources, computed within the WKB
approximation, have been studied within the contexts of Two Higgs
Doublet models~\cite{Fromme} and the SM with higher
dimensional operators~\cite{Huber:2006ri}.

\subsubsection{The baryon asymmetry}
Here, we compute the baryon asymmetry.  As argued above, we expect
that the leading BMSSM effects will enter through the CP-violating
sources.  We neglect BMSSM effects arising in the various transport
coefficients and diffusion constants that enter the Boltzmann
equations, following the general MSSM setup described in
Ref.~\cite{Chung:2009qs}.  In addition, we make the further assumption
of chemical equilibrium between particles and their superpartners,
valid when gauginos have masses $M_i \lesssim 1$ TeV.  We include
transport coefficients for bottom and tau Yukawa interactions,
recently shown to play an important role in MSSM
baryogenesis~\cite{Chung:2009cb}.

\begin{figure}[!t]
\begin{center}
\mbox{\hspace*{0cm}\epsfig{file=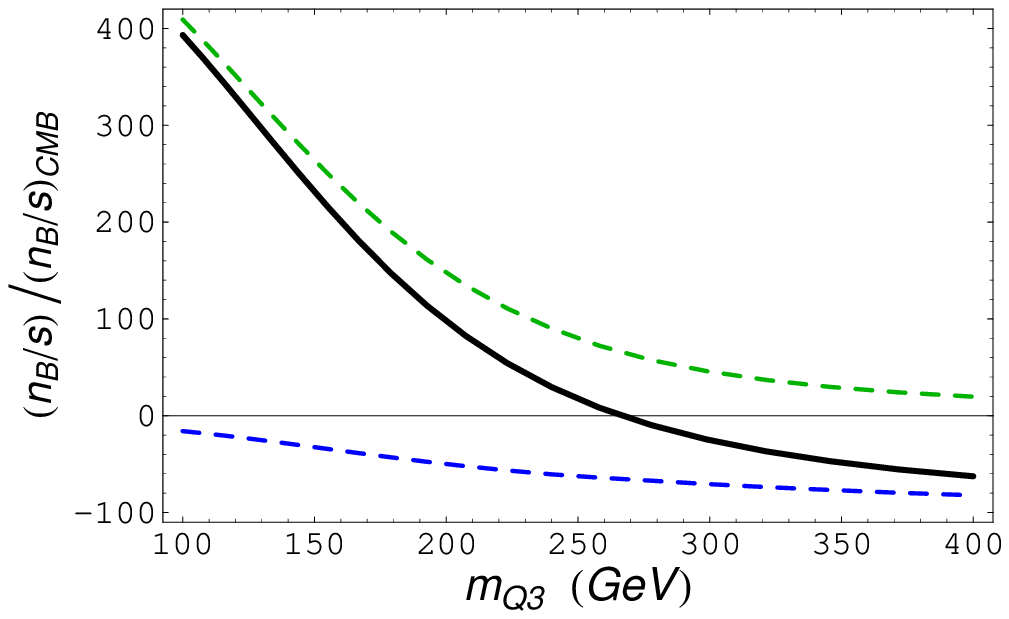,height=5cm}}
\mbox{\hspace*{0cm}\epsfig{file=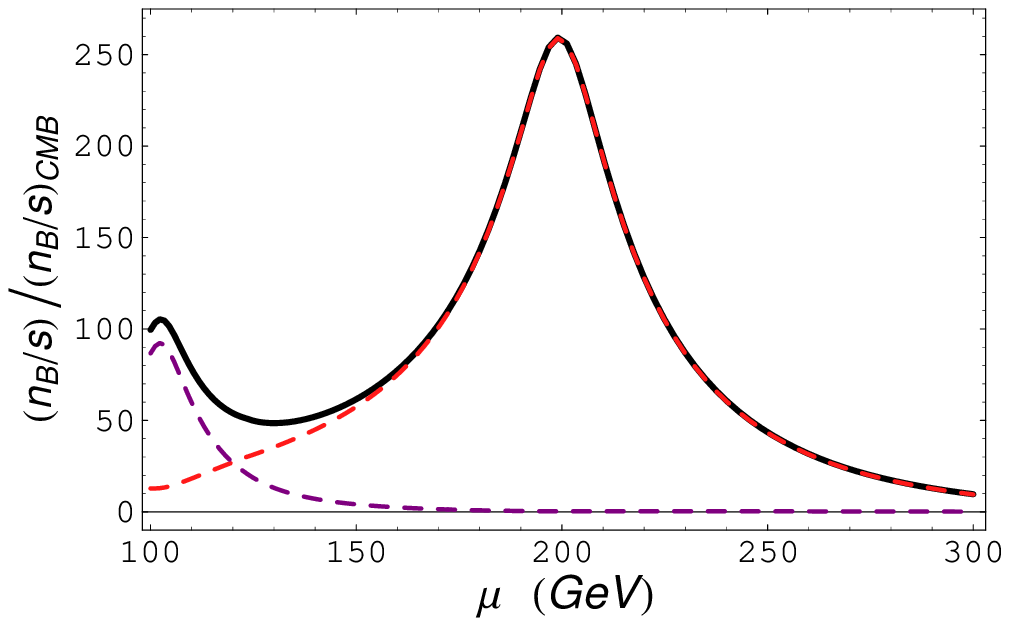,height=5cm}}
\end{center}
\caption{\it\small The baryon asymmetry generated by $\dot\theta$-terms in
  CP-violating sources in units of the observed BAU.  Left panel: top
  source (blue, dashed), stop source (green, dashed), and their sum
  (black, solid) as a function of the soft LH stop mass parameter.  Right
  panel: higgsino-bino source (purple, dashed), higgsino-wino source (red,
  dashed), and their sum (black, solid) as a function of $\mu$. Other
  relevant parameters are specified in Table~\ref{tab:params}. }
\label{fig:bauplots}
\end{figure}

The BMSSM CP-violating sources can have a large impact on baryon
number generation, illustrated in Fig.~\ref{fig:bauplots}.  Here, we
plot $n_B/s$ (the baryon-to-entropy-density ratio), normalized to the
observed value $n_B/s \simeq 9 \times 10^{-11}$~\cite{Dunkley:2008ie},
for maximal $\Delta\theta$; {\it i.e.}, the vertical axis approximately
shows $1/\Delta\theta$ needed to give the observed BAU. In order to
highlight the novel effects of the BMSSM, we take $\Delta\beta = 0$
and neglect all MSSM phases ($\phi_i = \phi_q = 0$).  The $\dot\theta$
contributions are suppressed in the large $\tan\beta$-limit; we take
$\tan\beta = 3$.  Other parameters are specified in
Table~\ref{tab:params}.

\begin{table}[b!]
\begin{tabular}{|@{\hspace{2mm}}c@{\hspace{2mm}}|@{\hspace{2mm}}c@{\hspace{2mm}}|@
{\hspace{2mm}}c@{\hspace{2mm}}|@{\hspace{2mm}}c@{\hspace{2mm}}|@{\hspace{2mm}}c@
{\hspace{2mm}}|@{\hspace{2mm}}c@{\hspace{2mm}}|@{\hspace{2mm}}c@{\hspace{2mm}}|@
{\hspace{2mm}}c@{\hspace{2mm}}||@
{\hspace{2mm}}c@{\hspace{2mm}}|@
{\hspace{2mm}}c@{\hspace{2mm}}|@
{\hspace{2mm}}c@{\hspace{2mm}}|}
\hline
$\mu$ & $M_1$ & $M_2$ & $M_3$ & $m_{Q_3}$ & $m_{\bar{u}_3}$ &
$m_{\bar{d}_3}$ & $m_{H_\pm}$ & $v_c$ & $T_c$ & $L_w^{-1}$ \\
\hline
400 & 100 & 200 & 500 & 300 & $\sqrt{-60^2}$ & 500 & 350 & 70 & 90 & 3 \\
\hline
\end{tabular}
\caption{\it\small The parameters used for Fig.~\ref{fig:bauplots}, in
  units of GeV.  In addition, we take $A_t =
A_b = 0$, $\tan\beta = 3$, and all other soft SUSY-breaking masses to be
heavier than 1 TeV.
 \label{tab:params}}
\end{table}

In the left panel of Fig.~\ref{fig:bauplots}, we show the BAU induced
via $\dot\theta$ contributions to the top (blue dashed) and stop
(green dashed) CP-violating sources, as a function the left-handed
stop mass parameter $m_{Q_3}$.  Since the right-handed stop is light,
the stop source is enhanced for smaller values of $m_{Q_3}$ due to the
resonance of the CP-violating source.  The stop source becomes
suppressed off-resonance, for large $m_{Q_3}$.  The top CP-violating
source does not depend on $m_{Q_3}$.  However, the top-driven
contribution to $n_B/s$ is suppressed at small values of $m_{Q_3}$ due
to (i) an enhanced stop contribution to the relaxation
rate~\cite{Lee:2004we}, and (ii) more charge equilibrating into
left-handed stops, rather than tops, reducing the fermionic charge
available for sphaleron conversion.

In the right panel of Fig.~\ref{fig:bauplots}, we show the BAU induced
via $\dot\theta$ contributions to the higgsino-wino (red dashed) and
higgsino-bino (purple dashed) CP-violating sources.  Both sources are
enhanced on-resonance when $\mu \approx M_{1,2}$.  In addition, we have
studied the bottom and sbottom CP-violating sources (not shown);
though similar to the top/stop contributions, they are suppressed by
$(m_b/m_t)^2$ and would be viable only for $\Delta\theta =
\mathcal{O}(1)$.

The total baryon asymmetry in the BMSSM is a combination of (i)
contributions shown in Fig.~\ref{fig:bauplots}, and (ii) standard MSSM
contributions induced through $\Delta\beta$ and the MSSM phases.  Our
key point is that the BMSSM contributions can be large, thereby
extending the window in parameter space that can provide successful
EWBG.  In the next section, we investigate the EDM constraints on this
scenario.

\section{Electric Dipole Moments}
\label{sec:edms}
\subsection{EDMs from BMSSM Phases}
EDM searches are sensitive to the same CP-violating phases that
generate the BAU, and consequently provide powerful constraints on the
EWBG mechanism. Currently, the most significant EDM bounds are for the
neutron and the thalium and mercury atoms
\cite{Regan:2002ta,Baker:2006ts, Griffith:2009zz}, given at 95\% C.L.:
\begin{subequations}
\label{eq:threeedms}
\beq
|\,d_n| &<& 3.5 \times 10^{-26} \; \; e \, \textrm{cm} \; ,\\
|\,d_{\textrm{Tl}}| &<& 1.1 \times 10^{-24}\; \; e \, \textrm{cm}\; , \\
|\,d_{\textrm{Hg}}| &<& 2.9 \times 10^{-29} \; \; e \, \textrm{cm}  \; .
\eeq
\end{subequations}
In many scenarios, including in our present work, the electron EDM
$d_e$ provides the dominant contribution to $d_{\textrm{Tl}}$, given
by $d_{\textrm{Tl}} \simeq - 585 \, d_e$.  Under this assumption, the
corresponding bound is $|\,d_e| < 1.9 \times 10^{-27} \, e$ cm [95\%
C.L.].

\begin{figure}[!t]
\begin{center}
\mbox{\hspace*{0cm}\epsfig{file=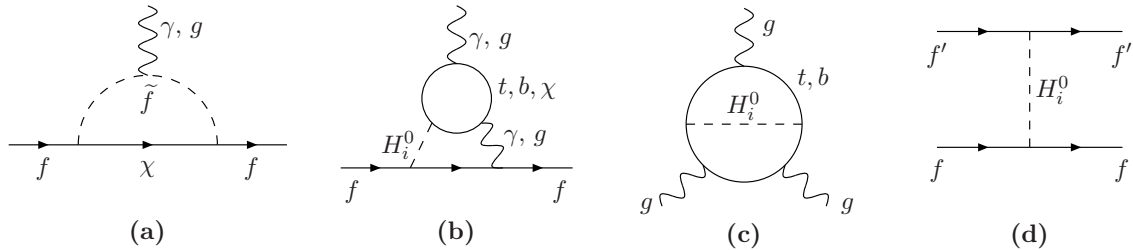,height=3.3cm}}
\end{center}
\caption{\it Examples of BMSSM contributions to CP-violating
  operators: (a) one-loop and (b) two-loop EDM and chromo-EDM, (c)
  Weinberg operator, and (d) four-fermion operator.
} \label{fig:feynman}
\end{figure}

CP violation in the BMSSM generates, below the weak scale, several
classes of CP-violating, non-renormalizable operators, which in turn
give rise to the above EDMs~\cite{Pospelov:2005pr}.  In
Fig.~\ref{fig:feynman}, we show examples of BMSSM contributions to
these operators.  At dimension five, there are EDM and chromo-EDM
operators, arising at one-loop order (Fig.~\ref{fig:feynman}a).
Two-loop contributions (Fig.~\ref{fig:feynman}b) become dominant when
first and second generation sfermions are heavy ($m_{\widetilde f}
\gtrsim 1$ TeV)\;\footnote{In the BMSSM, one must not push the first
  and second generation sfermion masses above the scale $M$, where the
  BMSSM ceases to be valid.  In principle, one could consider a
  modified version of the BMSSM in which these sfermions are
  integrated out along with the physics responsible for the BMSSM
  operators.}.  At dimension six, there are the Weinberg operator
(Fig.~\ref{fig:feynman}c) ~\cite{Weinberg:1989dx} and four-fermion
operators (Fig.~\ref{fig:feynman}d).  Novel BMSSM contributions to
these operators arise from (i) explicit factors of $\epsilon_{1i}$ in
the mass terms for sfermions and higgsinos
[Eqs.~(\ref{eq:mstop}-\ref{eq:neutralinomass})], (ii) tree-level
scalar-pseudoscalar neutral Higgs mixing [Eq.~(\ref{eq:m2Hpartdiag})],
and (iii) the complex Higgs vev, also arising at tree-level
[Eq.~(\ref{eq:imagmin1})].  These contributions give rise to
irreducible EDMs that cannot be universally suppressed without also
destroying the viability of EWBG (barring fine-tuned cancellations).

We evaluate these EDMs following Ref.~\cite{Ellis:2008zy}.  This
treatment, utilized for the MSSM with a loop-induced CP-violating
Higgs sector, includes all four classes of contributions shown in
Fig.~\ref{fig:feynman} and allows for the inclusion of the BMSSM
effects described above.  However, our EDM results are subject to two
main theoretical uncertainties.  First, the two-loop EDM and
chromo-EDM contributions are incomplete; the remaining known MSSM
contributions~\cite{Chang:1998uc,Pilaftsis:1999td,Li:2008kz} have not
been generalized to include scalar-pseudoscalar Higgs mixing and
cannot be easily adapted to the BMSSM.  Second, there exist
$\mathcal{O}(1)$ uncertainties in the hadronic inputs needed for the
evaluation of $d_{\textrm{Hg}}$ and $d_n$.  For the neutron, in
particular, there exist three different methods for computing $d_n$,
each sensitive to a different linear combination of CP-violating
operators.  Below, we show only neutron EDM bounds computed using the
more recent ``QCD Sum Rules'' method, which is sensitive to first
generation EDM, chromo-EDM, and Weinberg
operators~\cite{Pospelov:2000bw}.

As we show below, the mercury EDM provides the strongest bound on EWBG
in the BMSSM.  (In contrast, EWBG in the MSSM is constrained by $d_e$
and $d_n$ and is largely insensitive to $d_{\textrm{Hg}}$.)  This
scenario will be decisively probed by the combination of future EDM
searches, which are expected to reach sensitivities of $10^{-28} \, e$
cm for $d_n$, $10^{-29} \, e$ cm for $d_e$, and $10^{-29} \, e$ cm for
the EDMs of the deuteron $(d_D)$ and proton~\cite{futureedms}.

\subsection{Constraints on BMSSM Baryogenesis}
In this section, we show how limits on EDMs constrain BMSSM
baryogenesis.  Although both MSSM and BMSSM phases, listed in
Table~\ref{tab:phases}, can impact both EWBG and EDMs, we choose to
highlight the BMSSM by setting all MSSM phases to zero. In this case,
CP violation is governed by the parameters $\epsilon_{1i}$ and
$\epsilon_{2i}$.

\begin{figure}[!t]
\begin{center}
\mbox{\hspace*{-0.5cm}\epsfig{file=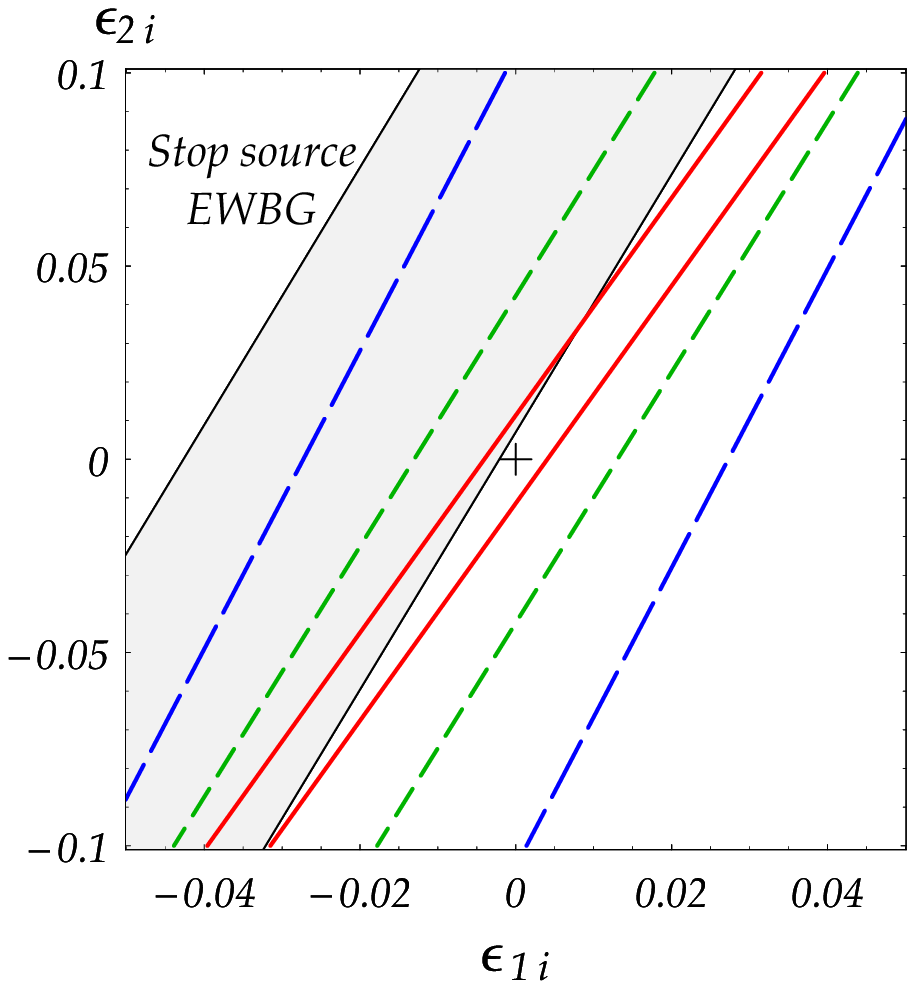,height=8cm}} \mbox{\hspace*{0cm}\epsfig{file=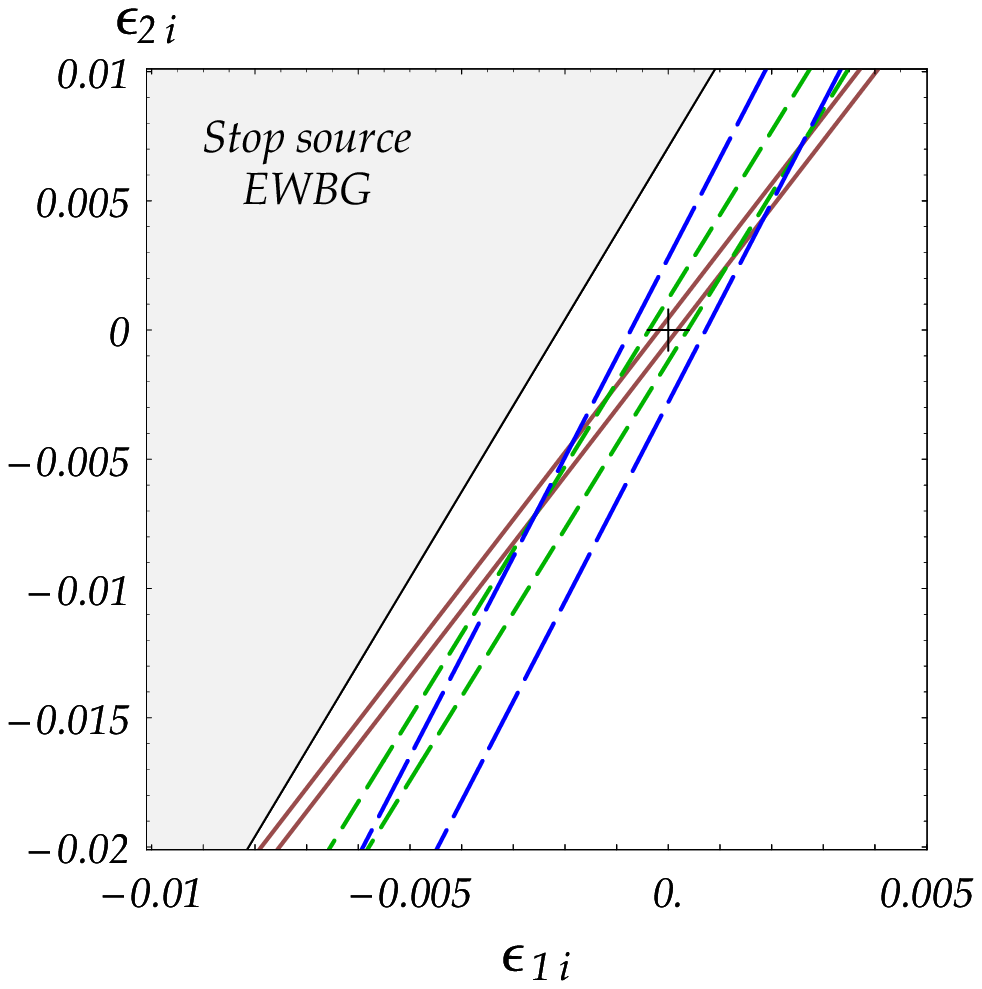,height=8cm}}
\end{center}
\caption{\it The viable region for stop-driven EWBG in the
  $\epsilon_{1i}$-$\epsilon_{2i}$ parameter space (in gray), and EDM
  constraints from $d_{\textrm{Hg}}$ (red solid), $d_{n}$ (green
  short-dash), $d_{e}$ (blue long-dash), and $d_D$ (brown solid).
  Left panel: current EDM constraints; Right panel: future EDM
  constraints: $|d_n| < 10^{-27} \; e$ cm, $|d_e| < 5\!\times \!
  10^{-29} \; e$ cm, $|d_D| < 10^{-28} \; e$ cm. (Note: zoomed-in
  scale on right.) Other
  relevant parameters are specified in Table~\ref{tab:params2}.
 \label{fig:edmbands}}
\end{figure}

\begin{figure}[!t]
\begin{center}
\mbox{\hspace*{-.5cm}\epsfig{file=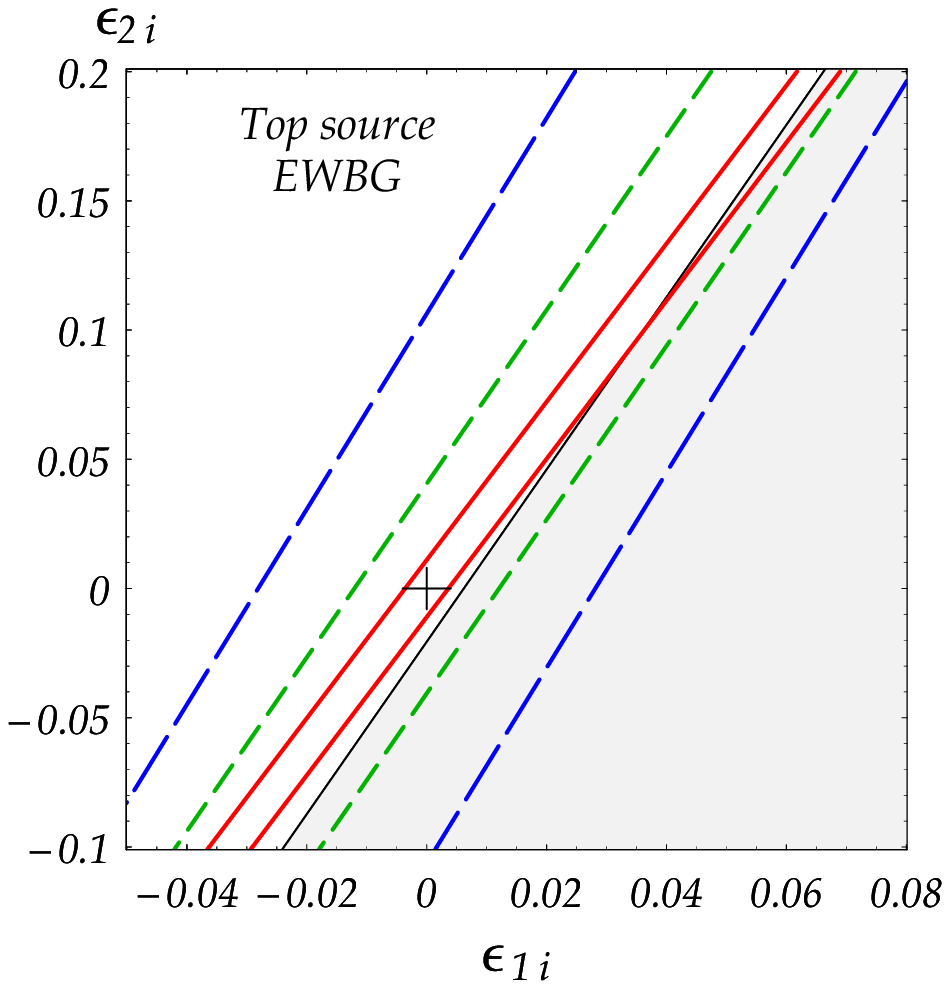,height=8cm}} \mbox{\hspace*{0cm}\epsfig{file=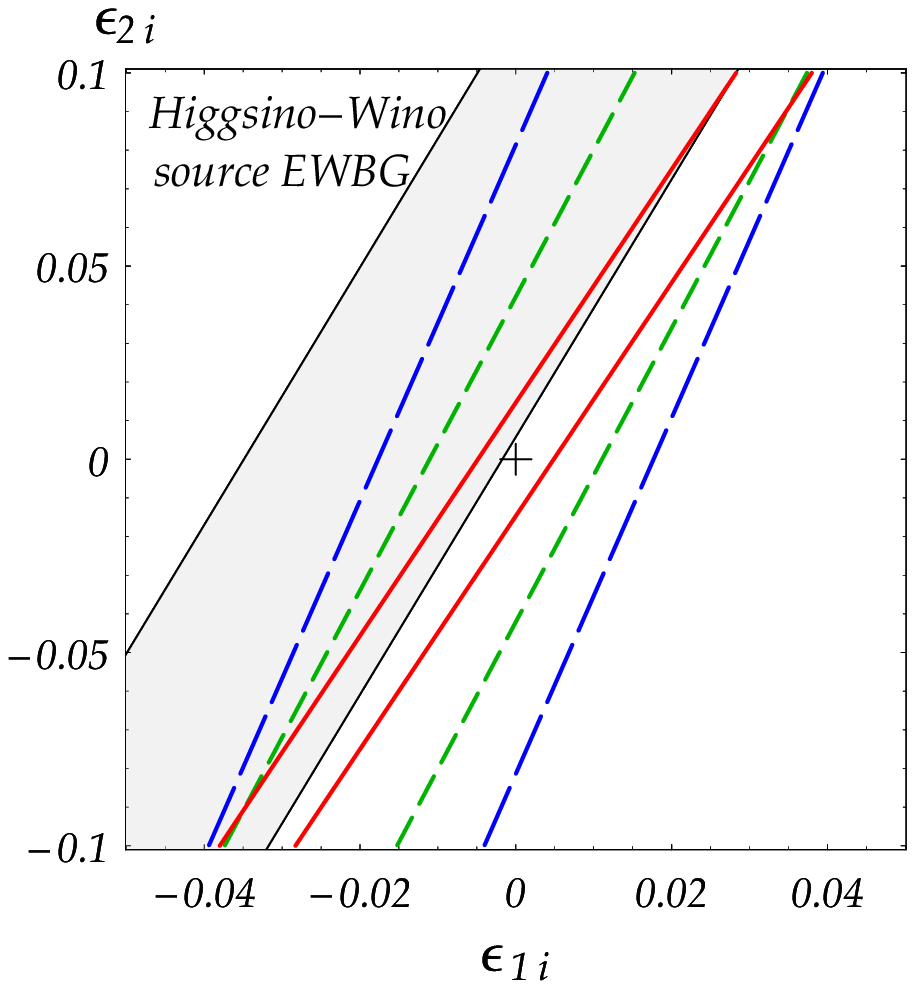,height=8cm}}
\end{center}
\caption{\it The viable region for top-driven (left) and higgsino-wino
  driven (right) EWBG in the $\epsilon_{1i}$-$\epsilon_{2i}$ parameter
  space (in gray), and EDM constraints from $d_{\textrm{Hg}}$ (red
  solid), $d_{n}$ (green short-dash), and $d_{e}$ (blue long-dash). Other
  relevant parameters are specified in Table~\ref{tab:params2}.
 \label{fig:edmbands2} }
\end{figure}

BMSSM baryogenesis can be driven by squark, quark, or higgsino
CP-violating sources, discussed in Sec.~\ref{sec:ewb}.  In
Figs.~\ref{fig:edmbands} and \ref{fig:edmbands2}, we illustrate how
current EDM constraints impact each of these EWBG scenarios, with
parameters given in Table~\ref{tab:params2}.  In each panel, the gray
bands show the region of the $\epsilon_{1i}$-$\epsilon_{2i}$ parameter
space consistent with generating the observed baryon asymmetry. BMSSM
baryogenesis is inconsistent with $\epsilon_{1i}=\epsilon_{2i}=0$
(shown by the cross), since clearly CP violation is required to
generate the BAU.  The width of this region corresponds to the following
range for the wall velocity $v_w$ and Higgs vev at the critical
temperature $v_c$ ({\it c.f.} Sec.~\ref{sec:ewb}):
\beq
0.01 < v_w < 0.4 \; , \qquad 70 \; \textrm{GeV} < v_c < 110 \; \textrm{GeV} \; .
\eeq
The edge of the EWBG region closest to the origin corresponds to the
minimum value of $(\epsilon_{1i} - \sin 2\beta \epsilon_{1i})$
consistent with the BAU, achieved when EWPT parameters fortuitously
maximize the CP-violating sources ($v_w \sim 0.03$ and $m_{U_3}^2$
such that the $v_c$ is largest).

In Fig.~\ref{fig:edmbands} we consider the stop-driven EWBG scenario.
In the left panel, the regions consistent with current mercury,
neutron and electron EDM constraints (95\% C.L.) lie between the red
(solid), green (short dash) and blue (long dash) curves, respectively.
Each EDM is consistent with $\epsilon_{1i} = \epsilon_{2i} = 0$.
Since the EWBG and EDM bands overlap, this scenario is viable in the
BMSSM.  (In the MSSM, it is not viable, since $m_{Q_3} \gg 1$
TeV~\cite{Carena:2008vj}.)  In the right panel, we illustrate how
future improvements in EDM sensitivities can exclude this scenario,
assuming null results.  Again, we show the same EWBG band, now zoomed
in.  The EDM bounds correspond to $|d_e| < 5 \!\times\! 10^{-29} \; e$
cm (blue long dash), $|d_n| < 10^{-27}\; e$ cm (green short dash), and
$|d_D| < 10^{-28}\; e$ cm (brown solid). EWBG is excluded within the
intersection of these bounds.  Furthermore, these are pessimistic
limits compared to actual expected future experimental sensitivities
(described above).  Of course, the alternative, that EDMs will be
discovered, is a much more exciting prospect.

\begin{table}[b!]
\begin{tabular}{|@{\hspace{2mm}}c@{\hspace{2mm}}|@{\hspace{2mm}}c@{\hspace{2mm}}|@{\hspace{2mm}}c@{\hspace{2mm}}|@
{\hspace{2mm}}c@{\hspace{2mm}}|@{\hspace{2mm}}c@{\hspace{2mm}}|@{\hspace{2mm}}c@
{\hspace{2mm}}|@{\hspace{2mm}}c@{\hspace{2mm}}|@{\hspace{2mm}}c@{\hspace{2mm}}|@
{\hspace{2mm}}c@{\hspace{2mm}}|}
\hline
EWBG scenario & $m_{Q_3}$ & $m_{\bar{u}_3}$ & $m_{\bar{d}_3}$ &
$m_{H_\pm}$ & $\mu$ & $M_1$ & $M_2$ & $M_3$ \\
\hline
squark-driven & 150 & $\sqrt{- 60^2}$ & 500 & 350 & 400 & 100 & 200 & 1000 \\
\hline
quark-driven & 500 & $\sqrt{- 60^2}$ & 500 & 350 & 400 & 100 & 200 & 1000 \\
\hline
wino-driven & 400 & $\sqrt{- 60^2}$ & 500 & 350 & 200 & 100 & 200 & 1000 \\
\hline
\end{tabular}
\caption{\it\small The parameters used for Figs.~\ref{fig:edmbands}
  and \ref{fig:edmbands2} in units of GeV.  We take $\tan\beta=3$,
  $\epsilon_{1r} = -0.05$, $\epsilon_{2r} = 0.05$, $A_f=0$, and $m_{\tilde
  f}^2 = (1\; \textrm{TeV})^2$ for all other sfermion
  masses-squared.
 \label{tab:params2}}
\end{table}

In Fig.~\ref{fig:edmbands2}, we consider EWBG scenarios driven by a
top CP-violating source (left) and higgsino-wino source (right).  As
above, viable EWBG can occur in the gray region, while the current EDM
constraints are shown by the colored bands, as in
Fig.~\ref{fig:edmbands2}.

Strikingly, the bands for EWBG and EDMs in Figs.~\ref{fig:edmbands}
and \ref{fig:edmbands2} appear to align in the
$\epsilon_{1i}$-$\epsilon_{2i}$ plane, signaling that they depend
largely on the same linear combination of $\epsilon_{1,2i}$.  The BAU
is proportional to $\Delta\theta$, the variation of the Higgs phase
across the wall, while the EDMs are dominated by contributions
proportional to $m^2_{hA}$ (the $h_0$-$A_0$ mixing parameter) or
$\theta$ (the vacuum Higgs phase). Indeed, all three parameters are
approximately proportional to $(2\epsilon_{1i} - \sin 2\beta \,
\epsilon_{2i})$.

The mercury EDM limit provides the strongest constraint on EWBG in the
BMSSM.  Here, the dominant contribution to $d_{\textrm{Hg}}$ is the
down quark chromo-EDM operator induced at two-loop
(Fig.~\ref{fig:feynman}b) with a top quark loop and light Higgs
exchange, proportional to the $h_0$-$A_0$ mixing parameter $m^2_{hA}$.
The subdominant contribution proportional to $m^2_{HA}$ is responsible
for the slight skew between the $d_{\textrm{Hg}}$ and EWBG bands in
Fig.~\ref{fig:edmbands}.

Lastly, we describe how EWBG and EDMs depend on the parameters $m_A$
and $\tan\beta$.  The consistency between EDMs and EWBG is generally
insensitive to $m_A$.  Larger values of $m_A$ suppress $\Delta\theta$
and the baryon asymmetry as $\sim 1/m_A^2$, requiring larger values of
$\epsilon_{1,2i}$, but the EDMs bounds are correspondingly weakened as
well.  On the other hand, for larger values of $\tan\beta$, the BMSSM
CP-violating sources are either constant or suppressed, while EDMs
become enhanced --- thus leading to stronger constraints on this
scenario.

\section{Conclusions}
\label{sec:conclude}
Adding dimension-five terms to the Higgs potential of the
supersymmetric standard model alleviates the fine-tuning problem
related to the LEP lower bound on the Higgs mass, and has interesting
consequences for the electroweak phase transition and dark matter.

Here, we have investigated the consequences of the new CP-violating
phases of these terms for supersymmetric baryogenesis and electric
dipole moments. Our main observations and conclusions are the
following:
\begin{enumerate}

\item The introduction of the $\epsilon_1$ and $\epsilon_2$ terms
  implies two new physical CP-violating phases.

\item Unlike the MSSM, the BMSSM allows for spontaneous baryogenesis,
  that is baryogenesis that is generated by a complex phase in the
  Higgs VEVs that is changing across the bubble wall.

\item In addition, several CP-violating sources that are ineffective
  in the MSSM, can become effective in the BMSSM, namely the
  stop and top sources.

\item It is possible to have successful baryogenesis with all the MSSM
  phases put to zero, and with either or both of the two new
  phases.

\item The EDM constraints can be satisfied if the new phases are of
  order $0.1$ or smaller. Successful baryogenesis requires,
  however, that they are not much smaller than the EDM upper bound.
  Thus, barring cancellations, BMSSM baryogenesis predicts that EDMs
  should be discovered if the experimental sensitivity improves by
  about an order of magnitude.

\end{enumerate}

\begin{appendix}
\section{Higgs phase vacua}\label{app:thtree}
In this appendix we discuss the vacuum value of the Higgs phase
$\theta$.  At tree level, the part of the Higgs potential which
depends on $\theta$ is
\beq
\Delta V_0=-v^2\sin2\beta\,{\rm Re}\left[e^{i\theta}
  \left(|b|+2|\epsilon_1|v^2e^{i\vartheta_1}\right)-
  e^{i(\vartheta_2+2\theta)}\frac{|\epsilon_2|v^2\sin2\beta}{2}\right]\;.
\eeq
At fixed values of $v$ and $\tan\beta$, the vacuum structure in the
$\theta$ direction is determined by a single complex parameter. To see this,
define the following quantities:
\beq\label{eq:effdef}
b_{\rm eff}&=&|b|+2|\epsilon_1|v^2e^{i\vartheta_1},\no\\
\vartheta_{\rm eff}&=&\arg[b_{\rm eff}],\no\\
|A|&=&|b_{\rm eff}|v^2\sin2\beta,\no\\
\kappa&=&-\left|\frac{\epsilon_2}{b_{\rm eff}}\right|v^2\sin2\beta
e^{i\left(\vartheta_2-2\vartheta_{\rm eff}\right)},\no\\
\tilde\theta&=&\theta+\vartheta_{\rm eff}\;.
\eeq
The potential and the minimum equations can be written as
\begin{subequations}\label{eq:thvaCProb}\beq
\Delta V_0(\tilde\theta)&=&-|A|\,{\rm Re}\left[e^{i\tilde\theta}
  +\frac{\kappa}{2}e^{2i\tilde\theta}\right],\\
\partial_{\tilde\theta}\Delta
V_0(\tilde\theta)&=&+|A|\,{\rm Im}\left[e^{i\tilde\theta}
  +\kappa e^{2i\tilde\theta}\right]=0,\label{eq:thvaCProb2}\\
\partial_{\tilde\theta}^2\Delta V_0(\tilde\theta)&=&+|A|\,
{\rm Re}\left[e^{i\tilde\theta}+2\kappa e^{2i\tilde\theta}\right]>0.\;\label{eq:thvaCProb3}
\eeq\end{subequations}
\eq{eq:thvaCProb} admits an analytic solution, depending only on
$\kappa$. The solution is obtained noting that \eq{eq:thvaCProb2} can
be cast as a quartic equation $\sum_{i=0}^4a_i\sin^i\tilde\theta=0$\,,
with coefficients $a_0={\rm Im}[\kappa]^2\,,a_1=2{\rm
  Im}[\kappa]\,,a_2=1-4|\kappa|^2\,,a_3=-4{\rm
  Im}[\kappa]\,,a_4=4|\kappa|^2$\,.  Out of the four roots of the
quartic equation, at most two represent a local minimum of the
potential.

In order for a double well potential to arise, the $\epsilon_2$ term
must be sizable in comparison with $b_{\rm eff}$. In particular, for
$|\kappa|<1/2$, only one minimum exists regardless of the phase of
$\kappa$\,. The condition on $\kappa$ can be cast as a condition on
the mass of the charged Higgs~\footnote{For $\kappa$ on the real axis,
  the emergence of an additional solution with $\theta\neq0$ implies
  spontaneous CP violation. This situation was discussed
  in~\cite{Ham:2009gu} for the BMSSM.} and on $\epsilon_2$,
\beq\label{eq:1thetaMin}
\left(\frac{m_{H_\pm}^2-m_W^2}{2v^2}\right)^2>3|\epsilon_2|^2-2\epsilon_{2r}
\left(\frac{m_{H_\pm}^2-m_W^2}{2v^2}\right)\;\;{\rm (condition\; for\;
  single\; minimum)}\;.  \eeq
The condition (\ref{eq:1thetaMin}) depends on neither $\tan\beta$ nor
$\epsilon_1$\,. For $|\epsilon_{2i,r}|\leq0.1$\,, it is fulfilled for
$m_{H_\pm}\gsim170$\,GeV. In case that $\epsilon_{2r}>0$\,, it is
enough to impose $m_{H_\pm}>130$\,GeV. In fact, violation of
\eq{eq:1thetaMin} necessarily implies
$m_{H_\pm}^2\sim\epsilon_2v^2$~\cite{Pomarol:1992bm}. This can be
understood in general from the Georgi-Pais
theorem~\cite{Georgi:1974au}. (In our case, nonrenormalizable
operators should be considered instead of quantum corrections to break
the CP-symmetry of the potential.) Such low values for $m_{H_\pm}$
(and hence also $m_A$) are, in general, in tension with both direct
and indirect experimental constraints. We did not pursue further the
analysis of this parameter regime, even though it may have interesting
consequences for EWBG via transitions between different phase
vacua~\cite{Comelli:1993ne}.

As a final comment, note that keeping the term of order $\epsilon_2^2$
in \eq{eq:1thetaMin} is required for small $m_{H_\pm}$\,, where
$(m_{H_\pm}^2-m_W^2)/2v^2$ becomes a small parameter
$\mathcal{O}(0.1)$\,. As is the case in several points in this work,
neglecting independent dimension six operators is still a consistent
procedure.

\section{Quantum Corrections}\label{app:QC}

Quantum corrections to the Higgs sector, in the presence of explicit
CP violation in the MSSM, were discussed in detail in
Refs.~\cite{Choi:2000wz,mssmCPv} for zero temperature and
Refs.~\cite{CPvbubmssm,Huber:1999sa} for finite temperature. Here we
present only the corrections that are directly relevant for the vev
phase in the BMSSM. We consider squark, chargino, neutralino and
scalar Higgs loops.

We write the potential as $V=V_0+\Delta V_1+\Delta V_T$, where $V_0$
is the tree-level, zero-temperature part, given in Eq. (\ref{eq:V0}),
$\Delta V_1$ is the zero-temperature one-loop part, and $\Delta V_T$
is the finite-temperature correction.  For $\Delta V_1$, one sums over
all particle species with field dependent masses:
\beq
\Delta
V_1=\sum_i\frac{n_im_i^4(\phi)}{64\pi^2}
\left(\ln\frac{m_i^2(\phi)}{Q^2}-\frac{3}{2}\right)\,,
\eeq
where $Q$ is the renormalization scale, which we choose as $Q=m_t$.
For $\Delta V_T$, we include the pressure and daisy terms:
\beq\label{eq:VTdaisy}
\Delta V_T=\sum_i\frac{n_iT^4}{2\pi^2}J
\left(\frac{m_i^2(\phi)}{T^2}\right)-\sum_{\rm i=sca}\frac{n_iT}{12\pi}
\left[m_i^3(\phi,T)-m_i^3(\phi)\right]\,,
\eeq
where the second sum includes scalars and longitudinal gauge bosons
and with the $J$ functions defined by
\beq
J_{B,F}(x)&=&\int_0^\infty dyy^2\ln\left(1\mp e^{-\sqrt{x+y^2}}\right)\;.
\eeq

Beginning with zero temperature, it is useful to first obtain an
analytical estimate of the contributions from stops, charginos and
neutralinos. We follow the procedure introduced in Section
\ref{sec:CPv}, and apply it to $V_0+\Delta V_1$. We find the corrected
expressions for $|b|$ and $\theta$,
\begin{subequations}\label{eq:bqcs}
\beq
|b|^2&=&\left[\frac{s_{2\beta}}{2}\left(m^2_{H^\pm}-m_W^2\right)
  +v^2\left(s_{2\beta}\epsilon_{2r}-2\epsilon_{1r}\right)
  -{\rm Re}\left[e^{i\theta}\delta\mathcal{V}'\right]\right]^2\no\\
&+&\left[v^2\left(s_{2\beta}\epsilon_{2i}-2\epsilon_{1i}\right)
  -{\rm Im}\left[e^{i\theta}\delta\mathcal{V}'\right]\right]^2,
\label{eq:bqc1}\\
\tan\theta&=&\frac{v^2\left(s_{2\beta}\epsilon_{2i}-2\epsilon_{1i}\right)
  -{\rm Im}\left[e^{i\theta}\delta\mathcal{V}'\right]}{\frac{s_{2\beta}}{2}
  \left(m^2_{H^\pm}-m_W^2\right)+v^2\left(s_{2\beta}\epsilon_{2r}
    -2\epsilon_{1r}\right)-{\rm Re}\left[e^{i\theta}\delta\mathcal{V}'\right]}\;.
\label{eq:bqc2}
\eeq
\end{subequations}
The quantity $\delta \mathcal{V}'(\propto \partial V/\partial\theta)$
encodes the various contributions:
\beq
\delta \mathcal{V}'&\approx&\delta_{\tilde
  t}\mathcal{V}'+\delta_{\tilde C}\mathcal{V}'+\delta_{\tilde
  N}\mathcal{V}',
\eeq
where, to $\mathcal{O}(\epsilon_1)$ and $\mathcal{O}(g^2)$, and
neglecting contributions to the charged Higgs mass arising from
diagrams involving chargino-neutralino and stop-sbottom loops, we have
\begin{subequations}\label{eq:qcanalytic}\beq
\delta_{\tilde
  t}\mathcal{V}'&\approx&\frac{3y_t^2}{16\pi^2}\mathcal{G}
\left(m_{\tilde t_2},m_{\tilde t_1}\right)\left[|A_t\mu|e^{i\phi_t}
  +2v^2|\epsilon_1|e^{i\vartheta_1}\left(c_\beta^2-s_{2\beta}
    \left|\frac{A_t}{\mu}\right|e^{i(\phi_t+\theta)}\right)\right]\\
\delta_{\tilde C}\mathcal{V}'&\approx&
-\frac{|\epsilon_1\mu^2|e^{i\vartheta_1}}{4\pi^2}\left(\ln\frac{|\mu|^2}{Q^2}-1\right)
+\frac{g^2|M_2\mu|e^{i\phi_2}}{8\pi^2}\,\mathcal{G}\left(|M_2|,|\mu|\right)\\
\delta_{\tilde
  N}\mathcal{V}'&\approx&-\frac{|\epsilon_1\mu^2|e^{i\vartheta_1}}{4\pi^2}\,
\left(\ln\frac{|\mu|^2}{Q^2}-1\right)+\frac{g^2|M_2\mu|e^{i\phi_2}}{16\pi^2}\,
\mathcal{G}(|M_2|,|\mu|)\no\\
&+&\frac{g'^2|M_1\mu|e^{i\phi_1}}{16\pi^2}\,\mathcal{G}(|M_1|,|\mu|)\;.
\eeq\end{subequations}
The loop function is
\beq
\mathcal{G}(m_a,m_b)&=&\frac{m^2_a
\ln\frac{m^2_a}{Q^2}-m^2_b\ln\frac{m^2_b}{Q^2}}{m^2_a-m^2_b}-1\;.\eeq

The conditions that the CP violation induced by the usual MSSM loop
corrections becomes comparable to the tree level nonrenormalizable
contribution  can be written as
\beq
\frac{\epsilon_1}{g^2/16\pi^2}&\lesssim&\frac{3M_2\mu}{2v^2}\;\Leftrightarrow\;
\frac{\epsilon_1}{0.05}\lesssim\frac{M_2\mu}{\left(600\,{\rm GeV}\right)^2},\no\\
\frac{\epsilon_1}{y_t^2/16\pi^2}&\lesssim&\frac{3A_t\mu}{2v^2}\;\Leftrightarrow\;
\frac{\epsilon_1}{0.05}\lesssim\frac{A_t\mu}{\left(400\,{\rm GeV}\right)^2}.\no
\eeq
We learn that the BMSSM terms can easily dominate. Furthermore, if
$\phi_{1,2,t}\ll1$, the radiative corrections are unimportant as they
mostly serve to slightly shift the value of $|b|$.

\eq{eq:qcanalytic} includes also a contribution from $\epsilon_1$
itself, arising via its appearance in the squark and sfermion mass
matrices. Here, the stop contribution is negligible compared to that
of charginos and neutralinos. This reflects the fact that $\epsilon_1$
corrects only the mass splitting of stops, while it enters the trace
of higgsino mass matrices. The $\epsilon_1$ loop correction tends to
cancel the tree level term. It becomes relevant if $\mu^2/v^2$ is
large enough to compensate for the loop suppression; in practice, the
term is significant for $\mu\gtrsim500$\,GeV.

Finally, a common feature of the zero-T corrections of
\eq{eq:qcanalytic} is that they depend only mildly on the vev. Hence,
the net effect of these terms in the vicinity of the origin of field
space is to globally shift the value of $\theta$. This implies that
the phase variation across the bubble wall, $\Delta\theta$, and
consequently the novel BMSSM contributions to the BAU, remain mostly
unchanged.

Proceeding to finite temperature, let us again obtain some analytical
understanding, beginning with the stop sector. We consider the
plausible limit where the heavier stop, $\tilde t_2$, is Boltzmann
suppressed, while the contribution of the lighter $\tilde t_1$ admits
a high-T expansion. Then,
\beq\label{eq:be2hT}
b_{\rm eff}(\phi,T)&\sim&|b|+2|\epsilon_1|\phi^2e^{i\vartheta_1}+\frac{3y_t^2T^2}{4m_{\tilde t_2}^2}
\left(|A_t\mu|e^{i\phi_t}+2|\epsilon_1|\phi^2c^2_\beta e^{i\vartheta_1}\right)\no\\
\epsilon_2(\phi,T)&\sim&\epsilon_2+\frac{3y_t^2T^2}{2m_{\tilde t_2}^2}
\frac{A_t\mu\epsilon_1}{|\mu|^2}\;.
\eeq
While there is no loop suppression in \eq{eq:be2hT}, the thermal
correction at the critical temperature is still down by a factor $\sim
T_c^2/m_{Q_3}^2$. Note also that the MSSM term $\propto
|A_t\mu|e^{i\phi_t}$ is field independent to 
$\mathcal{O}\left(T_c^2\phi^2/m_{Q_3}^4\right)$, such that its
contribution to the variation of $\theta$ across the bubble wall is
suppressed.  Using \eq{eq:be2hT}, we can compare the tree level effect
of the non-renormalizable BMSSM terms with the leading thermal
correction of the MSSM. Focusing on the phase variation along the
wall,
\beq\label{eq:bvsmssmfT}
\frac{\delta_{\rm BMSSM}}{\delta_{\rm
    MSSM}}\Big|_{_{T>0}}\sim0.5\left(\frac{\epsilon_1}{0.1}\right)
\left(\frac{m_{Q_3}^2}{A_t\mu}\right)\left(\frac{T_c}{100\,GeV}\right)^{-2}
\left(\frac{m_{Q_3}}{200\,GeV}\right)^2\;.
\eeq

In the bulk of this paper, we isolate the novel BMSSM effects by using
small or vanishing values of $A_t$, as well as vanishing $\phi_t$.
Hence only the term proportional to $\epsilon_1\phi^2c_\beta^2$ in
\eq{eq:be2hT} remains. Since this term is field dependent, it does not
affect the determination of $\theta$ in the symmetric outskirts of the
bubble wall, where $\phi\ll T_c$. Since it is doubly-$\tan\beta$
suppressed, it can typically be neglected with regard to the variation
of $\theta$ along the wall, even for $m_{Q_3}\sim T_c$.

Moving on to the Higgs and higgsinos, we find that Higgs-Higgs and
Higgs-higgsino interactions lead to a non-negligible shift in the
finite-temperature Higgs phase $\theta(z)$ at
$\mathcal{O}(\epsilon_1)$.  This effect is easy to understand by
considering a high temperature regime $T> m$, where $m$ denotes any
mass parameter in the Higgs and (weak) gaugino sectors. (Of course, at
such high $T>T_c$ the universe is globally symmetric. Nevertheless, it
is illuminating to provisionally pursue this line of argument.) In
this temperature regime and near the origin of field space, the
one-loop thermal potential due to Higgs and higgsino particles can be
expanded,
\beq\label{eq:VthighT}
V_{T}^{\rm H,\tilde H}\;&\sim& \;\frac{T^2}{24}\,{\rm Tr}M_{H_0}^2
+\frac{T^2}{12}\,{\rm Tr}M_{H_\pm}^2
+\frac{T^2}{24}\,{\rm Tr}{\bf M}_{\widetilde N}^\dag{\bf M}_{\widetilde N}
+\frac{T^2}{12}\,{\rm Tr}\mathbf X^\dag \mathbf X \no\\
\;&\supset&\; -\frac{3|\epsilon_1|T^2\phi^2s_{2\beta}}{2}
\cos(\theta+\vartheta_1)\;,
\eeq
The phase-dependent piece of the potential
becomes
\beq
V_T(\theta)\sim-\phi^2s_{2\beta}\left[|b|\cos\theta
+2|\epsilon_1|\left(\phi^2+\frac{3T^2}{4}\right)\cos(\theta+\vartheta_1)
-\frac{|\epsilon_2|\phi^2s_{2\beta}}{2}\cos(2\theta+\vartheta_2)\right]. \;
\eeq
In the high-T approximation and in the $\cot\beta \gg
|\epsilon_1|v^2/m_A^2$ regime, setting $T=T_c$, the values of $\theta$
in the symmetric $(s)$ and broken $(b)$ phases are
\beq\label{eq:thsapp}
\theta_{s} = \lim_{z \to -\infty} \theta(z)
&\sim&-\arctan\frac{3|\epsilon_1|T_c^2\sin\vartheta_1}
{2|b|+3|\epsilon_1|T_c^2\cos\vartheta_1}
\sim-\frac{3T_c^2\epsilon_{1i}}{m_{H_\pm}^2s_{2\beta}},\\
\label{eq:thTapp}
\theta_{b} = \lim_{z \to +\infty} \theta(z)  &\sim& \theta_s + \Delta\theta \;.
\eeq
In other words, the dominant effect is to shift $\theta(z)$ by
$\theta_s$ over the tree-level result [Eq.~\eqref{eq:thetaz}];
however, the relative phase $(\theta_b\!-\!\theta_s)$ that governs
EWBG remains unchanged from its tree-level value of $\Delta\theta$.

We should make the following comment, regarding the calculation of
$\theta_s$ in \eq{eq:thsapp}. Any dependence of the potential on
$\theta$ must be proportional to $\phi^2s_{2\beta}$. Thus, in the
truly symmetric region where the vev vanishes, the value of $\theta$
is not well defined. The quantity we denote by $\theta_s$ corresponds
to the value of the CP-violating phase in the almost symmetric regime,
where the vev is finite, but smaller than any other mass scale in the
problem. The non-zero value of $\theta_s$ indicates that, in the
complex plane, the origin can be approached from different directions.
Since $\theta$ cannot affect the dynamics in the symmetric regime, we
may extend the definition of $\theta_s$ to $z\to-\infty$. Having
clarified this point, it is important to keep in mind that in the
region of $\phi < g T$ there are additional non-perturbative
corrections to the scalar potential, which we do not consider here
beyond the daisy resummation introduced in \eq{eq:VTdaisy}. We note
that the daisy corrections, which include $\theta$-dependent terms in
the scalar self-energies, do not significantly affect our results for
$\theta_s$.
  
\begin{figure}[!t]
\begin{center}
\mbox{\hspace*{0cm}\epsfig{file=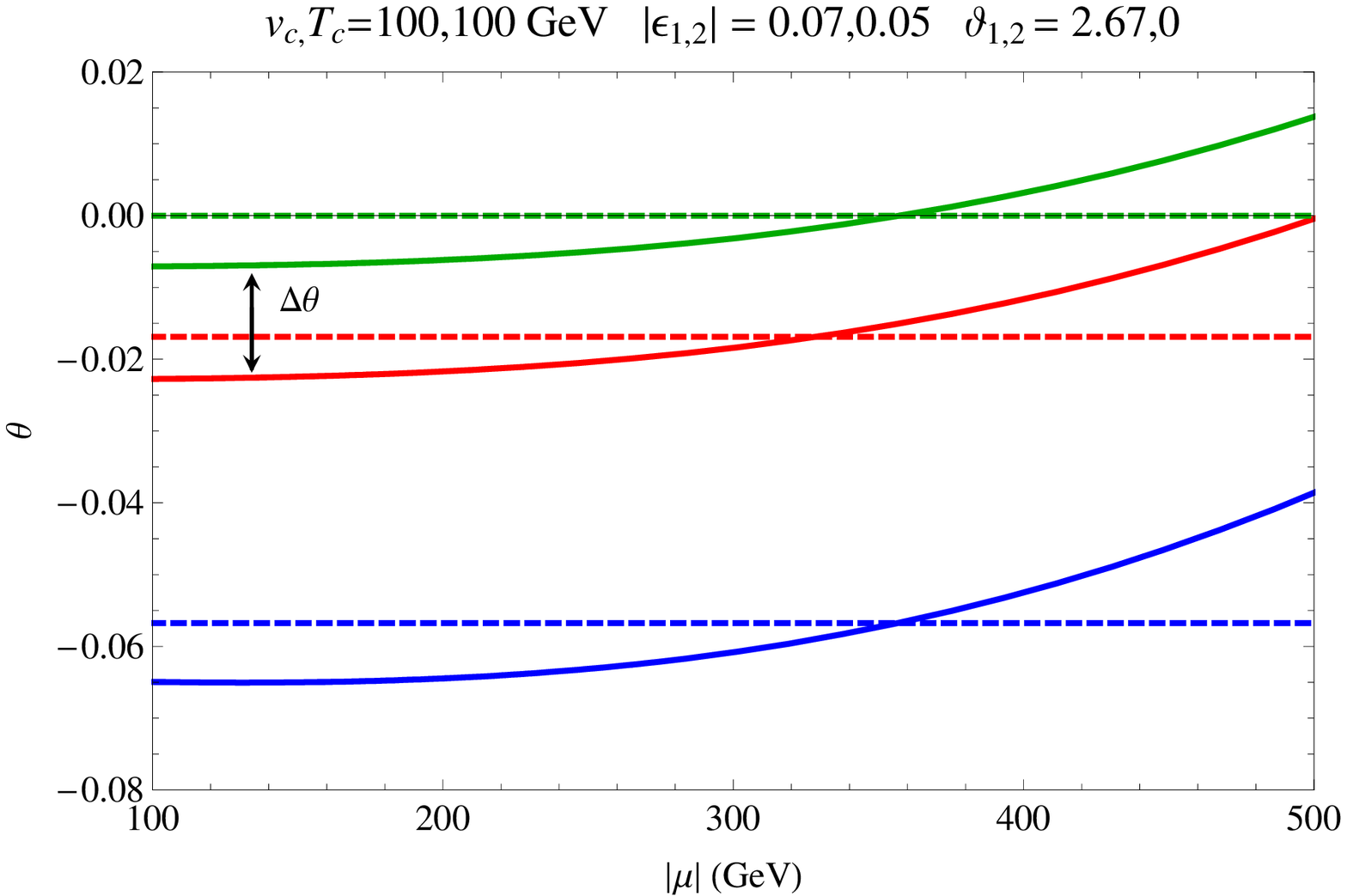,height=5.25cm}}
\mbox{\hspace*{0cm}\epsfig{file=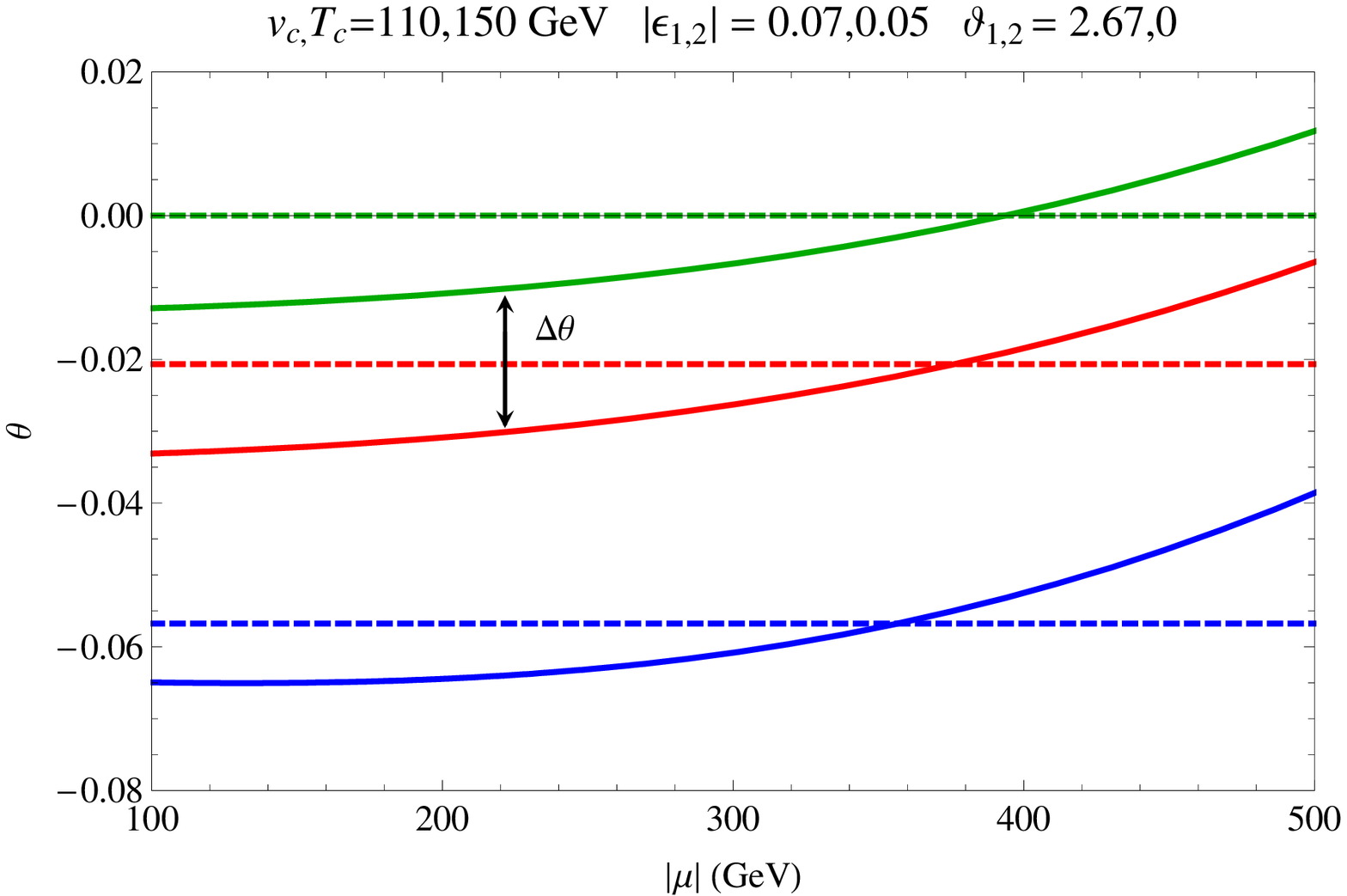,height=5.25cm}}
\end{center}
\caption{\it The vev phase, including quantum and thermal
  corrections. Unspecified parameters are as in
  Table~\ref{tab:params}. Left: $T_c=100$\,GeV, $v_c=100$\,GeV. Right:
  $T_c=150$\,GeV, $v_c=110$\,GeV. Green, red and blue curves
  correspond to $\theta_s$, $\theta_b=\theta_s+\Delta\theta$ and the
  zero-T value of $\theta$, respectively. Dashed (solid) curves denote
  tree level (one loop) results.  
} \label{fig:thetaQCMu}
\end{figure}

In \fig{fig:thetaQCMu} we illustrate the role of quantum and thermal
effects in dictating the complex vev phase at zero and finite
temperature. We go beyond the approximation of
\eqs{eq:thsapp}{eq:thTapp} by using the full one-loop thermal
potential, instead of the high-T expansion given in \eq{eq:VthighT}.
We solve for the asymptotic values of the complex phase by minimizing
the potential at the critical vev and near the origin of field space,
setting $T=T_c$. Since the $\epsilon_1$ loop contribution depends
mainly on $\mu$, we use $\mu$ as an independent variable. To emphasize
the role of the T-dependent terms, we fix all parameters and repeat
the plot in two panels, once for $T_c=v_c=100$ and once for
$T_c=150\,,\,v_c=110$\,GeV. We minimize the potential numerically,
accounting also for the zero- and finite-temperature effects of the
Higgs scalars.  Two main features which were mentioned above are worth
pointing out.  First, the onset of the zero temperature effect appears
at large $\mu\sim 400$\,GeV. Second, while quantum corrections shift
$\theta_s$ sizably, the value of $\Delta\theta$, the phase difference
between the broken and symmetric domains, is much less affected.

\end{appendix}

\begin{acknowledgments}
We thank J.R. Espinosa, T. Konstandin, D. Morrissey, A. Riotto and O. Vitells for helpful discussions.
The work of Y.N. is supported by the Israel Science Foundation (ISF)
under grant No.~377/07, by the German-Israeli foundation for
scientific research and development (GIF), and by the United
States-Israel Binational Science Foundation (BSF), Jerusalem,
Israel.

\end{acknowledgments}

\end{document}